\documentclass{article}
\usepackage{pgfplots}
\usepackage{lipsum}
\usepackage{caption}
\usepackage{subcaption}
\usepackage{authblk}
\usepackage[english]{babel}
\usepackage{amsfonts}
\usepackage{amsmath}
\usepackage[top=2cm, bottom=2cm, left=1.5cm, right=1.5cm]{geometry}
\usepackage{fancyhdr}
\usepackage{multicol}
\usepackage{bbold}
\usepackage{epsfig}
\usepackage{graphics}
\usepackage{epsfig}
\usepackage{lipsum}
\usepackage{fancyhdr}
\usepackage{booktabs}
\usepackage{amssymb}
\usepackage{wrapfig}
\usepackage[export]{adjustbox}
\usepackage{rotating}
\usepackage{amsmath}

\usepackage{hyperref}
\usepackage{cleveref}

\usepackage{cite}
\renewenvironment{abstract}{
	
	\hfill\begin{minipage}{0.95\textwidth}
		\rule{\textwidth}{1pt}}
	{\par\noindent\rule{\textwidth}{1pt}\end{minipage}}

\makeatletter
\usepackage{orcidlink}
\makeatother
\begin{document}

\bigskip \thispagestyle{empty}

\title{\textbf{Dynamic Evolution of Quantum Fisher and Skew Information under Decoherence in Three-Qubit X-States}}
	\author[1]{\textbf{A. Naimy}}
	\author[1,2]{\textbf{A. Slaoui\normalsize\orcidlink{0000-0002-5284-3240}}{\footnote {Corresponding author: {\sf abdallah.slaoui@fsr.um5.ac.ma}}}}
 \author[3]{\textbf{A. Ali\normalsize\orcidlink{0000-0001-9243-417X}}}
\author[1,2]{\textbf{H. El Hadfi}}
\author[1,2]{\textbf{R. Ahl Laamara\normalsize\orcidlink{0000-0002-8254-9085}}}
 \author[3]{\textbf{ S. Al-Kuwari}}
\affil[1]{\small LPHE-Modeling and Simulation, Faculty of Sciences, Mohammed V University in Rabat, Rabat, Morocco.}
	\affil[2]{\small Centre of Physics and Mathematics, CPM, Faculty of Sciences, Mohammed V University in Rabat, Rabat, Morocco.}
 \affil[3]{\small Qatar Centre for Quantum Computing, College of Science and Engineering, Hamad Bin Khalifa University, Doha, Qatar.}
	%
	%
	\maketitle
\begin{center}
    \textbf{Abstract}
\end{center}

\begin{abstract}
Quantum metrology leverages quantum effects such as squeezing, entanglement, and other quantum correlations to boost precision in parameter estimation by saturating quantum Cramer Rao bound, which can be achieved by optimizing quantum Fisher information or Wigner-Yanase skew information. This work provides analytical expressions for quantum Fisher and skew information in a general three-qubit X-state and examines their evolution under phase damping, depolarization, and phase-flip decoherence channels. To illustrate the validity of our method, we investigate their dynamics for a three-qubit Greenberger-Horne-Zeilinger (GHZ) state subjected to various memoryless decoherence channels. Closed-form expressions for QFI and SQI are derived for each channel. By comparing these metrics with the entanglement measure of concurrence, we demonstrate the impact of decoherence on measurement precision for quantum metrology. Our results indicate that phase damping and phase-flip channels generally allow for better parameter estimation compared to depolarization. This study provides insights into the optimal selection of noise channels for enhancing precision in quantum metrological tasks involving multi-qubit entangled states.
\end{abstract}


\section{Introduction}
A hallmark of quantum mechanics is entanglement, a phenomenon in which quantum systems, once they interact, become inextricably linked. This results in correlations that exceed classical expectations and persist even after spatial separation \cite{Wootters1998,Horodecki2009}. Importantly, this nonlocality does not violate the principles of relativity \cite{Einstein1935,Bell1964}. Additionally, quantum coherence—encompassing not only quantum superposition but also various types of non-classical correlations—serves as a fundamental resource underpinning a range of quantum technological applications, such as quantum simulation \cite{Georgescu2014,Kirdi2023,Slaoui2024}, error correction \cite{Shor1995}, secure quantum communication \cite{Gisin2002}, quantum cryptography \cite{Bennett2014}, and quantum metrology (QM) \cite{Giovannetti2011,Pezze2018}.

In quantum estimation theory, measurement precision is characterized by two distinctive scaling regimes: the standard quantum limit (SQL), which scales as $1/\sqrt{N}$ and is achieved through measurements with uncorrelated particles, and the Heisenberg limit (HL), which scales as $1/N$ and is attainable through various quantum-enhanced methodologies. The HL demonstrates significant metrological advantages over the SQL in quantum parameter estimation protocols. Experimental investigations have shown that utilizing $N$ correlated particles enables high-precision measurements \cite{Huang2016,Riedel2010}, with multi-qubit probe states outperforming single-qubit implementations in parameter estimation scenarios \cite{Hu2020}. The fundamental advantage of QM over classical approaches lies in its ability to exploit quantum effects such as superposition, squeezing, and quantum entanglement to achieve enhanced parameter estimation precision \cite{Giovannetti2004,SlaouiB2023,Dorner2012}. The achievable precision, $\delta\varphi$, is fundamentally bounded by the quantum Cramér-Rao inequality \cite{Braunstein1994}; $\delta\varphi \geq \frac{1}{\sqrt{n\mathcal{F}}}$, where $\varphi$ denotes the parameter being estimated, $\mathcal{F}$ represents the quantum Fisher information (QFI), and $n$ quantifies the measurement duration. As a result, the theoretical precision limit is determined by the quantum Fisher information \cite{Zheng2016,Wang2016}, making its optimization a central goal in quantum metrological protocols.\par

QFI framework offers a powerful lens into quantum simulation methodologies and high-precision quantum sensor design, enabling the detection and characterization of intricate quantum phenomena, such as superposition states and critical quantum dynamics. By quantifying the sensitivity of quantum states to parameter variations, QFI provides a fundamental limit on the precision of quantum metrology \cite{AbouelkhirS2023,Abdellaoui2024}. Furthermore, the Wigner-Yanase skew information, which quantifies the non-commutativity between a self-adjoint operator $K$ and a quantum state $\varrho$, plays a pivotal role in quantum estimation theory and information content characterization. This measure has proven particularly useful in quantifying non-classical correlations in bipartite and tripartite quantum systems, offering insights into the resourcefulness of quantum states for various information processing tasks \cite{Luo2004,Luo2012}. From a mathematical standpoint, several analytical techniques are available for calculating QFI in QM. A common approach involves diagonalizing the density matrix to utilize its eigenvalues and eigenvectors. Another widely used method relies on the symmetric logarithmic derivative (SLD), a Hermitian operator associated with the derivative of the density matrix with respect to the parameter of interest. This method is particularly advantageous for pure states and certain types of mixed states \cite{AbouelkhirH2023,Rath2021}. Furthermore, for specific quantum states and parameterizations, analytical QFI expressions can be obtained using tools from differential geometry or algebraic methods. While these techniques are highly effective, they often involve intricate mathematical computations and may not be practical for complex quantum systems.\par

In this direction, this paper aims to develop an analytical method to evaluate the QFI and SI in three-qubit systems. These quantities provide valuable insights into the quantum properties of a system, such as its sensitivity to parameter changes and its degree of coherence. We derive a general analytical expression for these quantities in three-qubit $X$-states, a class of states with a simplified density matrix structure. To illustrate the applicability of our method, we apply it to Werner-Greenberger-Horne-Zeilinger states, a family of mixed states that interpolate between a maximally mixed state and a pure GHZ-state \cite{Greenberger1990}. These states enable the testing of profound aspects of quantum theory while offering practical applications in the transmission and processing of quantum information. This article is structured as outlined below: in section two we present some fundamental properties of X states with three qubits and explain the method for calculating quantum Fisher information and skew information for these states. The exact formulations of these two values are given in Section three for the phase-damping, the depolarizing, and the phase-flip channels, among other decoherence channels. In the discussion , we explore the evolution of quantum Fisher and skew information for GHZ three-qubit states under noisy conditions. We also compare these results with Wootters concurrence in three different types of decoherence channels and present the numerical outcomes. Finally, we conclude this study in the last paragraph.

\section{Formulation of quantum Fisher and Skew information for a three-qubit X state}
\subsection{Three-qubit $X$-state}
The unique feature of three-qubit X-states lies in their capacity to facilitate analytical calculations, thus enhancing understanding and applications. These states have already proven effective in entanglement \cite{Yu2004,Sun2017} and quantum discord analyses \cite{Fanchini2010,Maziero2010}. Their density matrix is characterized by non-zero elements confined to diagonal and anti-diagonal entries, resembling the letter X, hence the term X-states. This class includes a diverse range of states—classical and non-classical, separable and non-separable—allowing focus on X-states without limiting the investigation. Extending this concept to multi-qubit systems applies to various states in quantum information processing, such as the Bell states \cite{Ghosh2001} and quasi-Werner states \cite{Werner1989} for two qubits, and Dicke states \cite{Guhne2009} for \(N\) qubits.
From now on, we will only consider the three-qubit $X$-states. Mathematically, the $X$-state 
    in the computational basis $\mathit{B}_0= \left\{ |000\rangle, |001\rangle, |010\rangle, |011\rangle, |100\rangle, |101\rangle, |110\rangle, |111\rangle \right\}$ is given by
\begin{equation}
		\varrho_{ABC}=\left(\begin{array}{cccccccc}
               \varrho_{11} & 0 & 0 & 0 & 0 & 0 & 0 & \varrho_{18}  \\
			0 & \varrho_{22} & 0 & 0 & 0 & 0 & \varrho_{72} & 0 \\
			0 & 0 & \varrho_{33} & 0 & 0 & \varrho_{36} & 0 & 0 \\
			0 & 0 & 0 & \varrho_{44} & \varrho_{45} & 0 & 0 & 0 \\
                0 & 0 & 0 & \varrho_{54} & \varrho_{55} & 0 & 0 & 0 \\
                0 & 0 & \varrho_{63} & 0 & 0 & \varrho_{66} & 0 & 0 \\
                0 & \varrho_{72} & 0 & 0 & 0 & 0 & \varrho_{77} & 0 \\
                \varrho_{81} & 0 & 0 & 0 & 0 & 0 & 0 & \varrho_{88} \\
		\end{array}\right), \label{matrix rho}
				\end{equation}

The parametrized Fano-Bloch representation of  $\varrho_{ABC}$ can be expressed as 
    \begin{equation}
        \varrho_{ABC} = \frac{1}{8} \sum_{\alpha \,\beta \, \gamma} \mathcal{T}_{\alpha \beta \gamma} \sigma_{\alpha} \oplus \sigma_{\beta} \oplus \sigma_{\gamma},
    \end{equation}
where, $\mathcal{T}_{\alpha \beta \gamma} = Tr( \varrho_{ABC} \, (\sigma_{\alpha} \oplus \sigma_{\beta} \oplus \sigma_{\gamma}))$ denotes the elements of the correlation matrix, with the constraints $\sum_{i=1}^4 \varrho_{ii}=1, \, \text{and}  \,\varrho_{ii}\,\varrho_{jj} \geq \lvert \varrho_{ij} \rvert^2 $. The Fano-Bloch representation can also be expressed with direct sum representation that reads 
    \begin{equation}
        \varrho_{ABC} =  \varrho_{ABC}^{(1)} \oplus \varrho_{ABC}^{(2)} \oplus\varrho_{ABC}^{(3)}\oplus\varrho_{ABC}^{(4)} \label{rho oplus}
    \end{equation}
    With these computational basis, $\mathit{B}_{1} =\left\{|000\rangle, |111\rangle \right\}, \mathit{B}_{2} = \left\{|001\rangle, |110\rangle \right\}, \mathit{B}_{3} = \left\{|010\rangle, |101\rangle \right\} \;\text{and}\;  \mathit{B}_{4} = \left\{|011\rangle, |100\rangle \right\}$, respectively, the submatrices $\varrho_{ABC}^{(1)}$, $\varrho_{ABC}^{(2)}$, $\varrho_{ABC}^{(3)}$ and $\varrho_{ABC}^{(4)}$ are defined as 
\begin{equation}
\varrho_{ABC}^{(1)}=\left(\begin{array}{ll}
	\varrho_{11} & \varrho_{18} \\
				\varrho_{81} & \varrho_{88}
			\end{array}\right), \hspace{0.2cm}  \varrho_{ABC}^{(2)}=\left(\begin{array}{ll}
				\varrho_{22} & \varrho_{27} \\
				\varrho_{72} & \varrho_{77}
			\end{array}\right), \hspace{0.2cm}
    \varrho_{ABC}^{(3)}=\left(\begin{array}{ll}
				\varrho_{33} & \varrho_{36} \\
				\varrho_{63} & \varrho_{66}
			\end{array}\right) \hspace{0.1cm} \text{and} \;\hspace{0.1cm}
    \varrho_{ABC}^{(4)}=\left(\begin{array}{ll}
				\varrho_{44} & \varrho_{45} \\
				\varrho_{54} & \varrho_{55}
			\end{array}\right).
		\end{equation}
 These density matrices $\varrho_{ABC}^{j}$ can be represented in the Lie algebra of the unitary group SU(2) using the Pauli matrices $\sigma_{k}^{B_j}$ (with $k= 1,2,3$) and the identity matrices $\mathbb{I}_{B_j}^j$ in the computational basis ${B_j}^j$. In this framework \cite{Bellorin2022}, these matrices $\varrho_{ABC}^{j}$ are expressed as
\begin{equation}
    \varrho_{ABC}^{j}=\,\frac{1}{2} \, (\omega_0\,\mathbb{I}_{B_j}^j+\sum_{k=1}^{3}\,\omega_k^j\,\sigma_k^{B_j})= \,\frac{1}{2} \sum_{\alpha= 0}^{3} \omega_\alpha^{j} \eta_\alpha^{j},
\end{equation}
    with $\eta_0^{j}=\mathbb{I}_{B_j}^j$ for $\alpha=0$, and $\eta_\alpha^{j}=\sigma_\alpha^{B_j}$ for $\alpha=k$. The generators $\eta_1$, $\eta_2$, $\eta_3$ and $\eta_4$, when expressed in the computational basis, are as follows
\begin{align}
\left\{
\begin{array}{ll}
\eta_0^1=|000\rangle \langle 000|+|111\rangle \langle 111|,\\
\eta_1^1=|000\rangle \langle 111|+|111\rangle \langle 000|,\\
\eta_2^1=i|111\rangle \langle 000|-i|000\rangle \langle 111|,\\
\eta_3^1=|000\rangle \langle 000|-|111\rangle \langle 111|,
\end{array} 
\right.
\hspace{2cm}
\left\{
\begin{array}{ll}
\eta_0^2=|001\rangle \langle 001|+|110\rangle \langle 110|,\\
\eta_1^2=|001\rangle \langle 110|+|110\rangle \langle 001|,\\
\eta_2^2=i|110\rangle \langle 001|-i|001\rangle \langle 110|,\\
\eta_3^2=|001\rangle \langle 001|-|110\rangle \langle 110|,
\end{array} 
\right. \\
\left\{
\begin{array}{ll}
\eta_0^3=|010\rangle \langle 010|+|101\rangle \langle 101|,\\
\eta_1^3=|010\rangle \langle 101|+|101\rangle \langle 010|,\\
\eta_2^3=i|101\rangle \langle 010|-i|010\rangle \langle 101|,\\
\eta_3^3=|010\rangle \langle 010|-|101\rangle \langle 101|,
\end{array} 
\right.
\hspace{2cm}
\left\{
\begin{array}{ll}
\eta_0^4=|011\rangle \langle 011|+|100\rangle \langle 100|,\\
\eta_1^4=|011\rangle \langle 011|+|100\rangle \langle 011|,\\
\eta_2^4=i|100\rangle \langle 011|-i|011\rangle \langle 100|,\\
\eta_3^4=|011\rangle \langle 011|-|100\rangle \langle 100|.
\end{array} 
\right.
\end{align}
To calculate the expressions for the quantum Fisher and skew information, it is evident that in this decomposition, the matrices $\varrho_{ABC}^{(1)}$, $\varrho_{ABC}^{(2)}$, $\varrho_{ABC}^{(3)}$ and $\varrho_{ABC}^{(4)}$ take on the form
\begin{equation}
    \varrho_{ABC}^{(1)}=\frac{1}{2} \sum_{\alpha=0}^{3} \omega_{\alpha}^1 \eta_{\alpha}^1,
    \hspace{0.5cm} 
     \varrho_{ABC}^{(2)}=\frac{1}{2} \sum_{\alpha=0}^{3} \omega_{\alpha}^2 \eta_{\alpha}^2,
    \hspace{0.5cm} 
     \varrho_{ABC}^{(3)}=\frac{1}{2} \sum_{\alpha=0}^{3} \omega_{\alpha}^3 \eta_{\alpha}^3
    \hspace{0.3cm} \text{and} \hspace{0.3cm}
     \varrho_{ABC}^{(4)}=\frac{1}{2} \sum_{\alpha=0}^{3} \omega_{\alpha}^4 \eta_{\alpha}^4, \label{rho in fbch rep}
\end{equation}
where $\omega_{\alpha}^1$, $\omega_{\alpha}^2$, $\omega_{\alpha}^3$, and $\omega_{\alpha}^4$ are provided in the following order 
\begin{equation}
    \begin{array}{cccc}
         & \omega_0^1=\frac{1}{4}\left( \mathcal{T}_{000}+\mathcal{T}_{033}+\mathcal{T}_{303}+\mathcal{T}_{330} \right), \hspace{1.5cm} \omega_0^2=\frac{1}{4}\left( \mathcal{T}_{000}-\mathcal{T}_{033}-\mathcal{T}_{303}+\mathcal{T}_{330} \right),\\
         
        & \omega_1^1=\frac{1}{4}\left( \mathcal{T}_{111}-\mathcal{T}_{122}-\mathcal{T}_{212}-\mathcal{T}_{221} \right),\hspace{1.5cm} \omega_1^2=\frac{1}{4}\left( \mathcal{T}_{111}+\mathcal{T}_{122}+\mathcal{T}_{212}-\mathcal{T}_{221} \right),\\
        
       & \omega_2^1=\frac{1}{4}\left(-\mathcal{T}_{112}-\mathcal{T}_{121}+\mathcal{T}_{211}-\mathcal{T}_{222} \right),\hspace{1.5cm} \omega_2^2=\frac{1}{4}\left( \mathcal{T}_{112}-\mathcal{T}_{121}-\mathcal{T}_{211}-\mathcal{T}_{222}\ \right),\\
       
       & \omega_3^1=\frac{1}{4}\left( \mathcal{T}_{003}+\mathcal{T}_{300}+\mathcal{T}_{333}+\mathcal{T}_{030} \right),\hspace{1.5cm} \omega_3^2=\frac{1}{4}\left(-\mathcal{T}_{003}+\mathcal{T}_{030}+\mathcal{T}_{300}-\mathcal{T}_{333} \right). \label{ommega 1,2}
    \end{array}
\end{equation}

\begin{equation}
    \begin{array}{cccc}
         & \omega_0^3=\frac{1}{4}\left( \mathcal{T}_{000}-\mathcal{T}_{033}+\mathcal{T}_{303}-\mathcal{T}_{330} \right),\hspace{1.5cm} \omega_0^4=\frac{1}{4}\left( \mathcal{T}_{000}+\mathcal{T}_{033}-\mathcal{T}_{303}-\mathcal{T}_{330} \right),\\
         
         & \omega_1^3=\frac{1}{4}\left( \mathcal{T}_{111}+\mathcal{T}_{122}-\mathcal{T}_{212}+\mathcal{T}_{221} \right),\hspace{1.5cm} \omega_1^4=\frac{1}{4}\left( \mathcal{T}_{111}+\mathcal{T}_{122}+\mathcal{T}_{212}+\mathcal{T}_{221} \right),\\
         
         & \omega_2^3=\frac{1}{4}\left(-\mathcal{T}_{112}+\mathcal{T}_{121}-\mathcal{T}_{211}-\mathcal{T}_{222}\ \right), \hspace{1.5cm} \omega_2^4=\frac{1}{4}\left( \mathcal{T}_{112}+\mathcal{T}_{121}-\mathcal{T}_{211}+\mathcal{T}_{222}\ \right),\\
         
         & \omega_3^3=\frac{1}{4}\left(\mathcal{T}_{003}-\mathcal{T}_{030}+\mathcal{T}_{300}-\mathcal{T}_{333} \right), \hspace{1.5cm} \omega_3^4=\frac{1}{4}\left(-\mathcal{T}_{003}-\mathcal{T}_{030}+\mathcal{T}_{300}+\mathcal{T}_{333} \right),  \label{ommega 3,4}
    \end{array}
\end{equation}

\subsection{Detailed expression of quantum Fisher information}

Assume $\varrho_\varphi$ is a quantum state influenced by an arbitrary parameter $\varphi$. The density matrix $\varrho_\varphi$ requires a sequence of quantum measurements $\{ \Pi_x \}$ to estimate value of $\varphi$. In classical metrology, Fisher information sets the minimum variance achievable by an unbiased estimator for the parameter being estimated, we define it as
\begin{equation}
    \texttt{F}_c(\varphi)=\int \mathsf{p}(x/\varphi)\left[ \frac{\partial \, ln\, \mathsf{p}(x/\varphi)}{\partial\varphi} \right]^2 dx. \label{Fc}
\end{equation}
In this context,  $\mathsf{p}(x/\varphi) = \mathrm{Tr}\left[\Pi_x \varrho_\varphi\right]$ denotes the probability of observing the outcome $x$ when the actual parameter value is $\varphi$. In accordance with classical statistical theory, the limiting precision of estimating $\varphi$ from $N$ measurements, represented by the variance $\text{Var}(\varphi)$, is limited below by the Fisher information $\texttt{F}_c(\varphi)$, this bound is defined by the Cramér-Rao formula, as expressed below
\begin{equation}
    \text{Var}(\varphi) \ge\frac{1}{\textit{N} \,\texttt{F}_c(\varphi)}.
\end{equation}
The maximum of classical Fisher information (Eq. \eqref{Fc}) across the set of possible measurements $\{\Pi_x\}$ is known as quantum Fisher information
\begin{equation}
    \mathbf{\textit{F}}(\varrho_\varphi)= \underset{\Pi_x}{\text{max}}\, \texttt{F}_c(\varphi)  ,
\end{equation}
we can now express the equation above in a more explicit form
\begin{equation}
    \mathbf{\textit{F}}(\varrho_{\varphi})= Tr(\varrho_{\varphi}L^{2}_\varphi)= Tr[(\partial_{\varphi} ,\varrho_{\varphi})L_\varphi],
\end{equation}
where the symmetrical logarithmic derivative (SLD) operator $L_\varphi$ \cite{Liu2016}, is explicitly found via the following relation.
\begin{equation}
    \frac{\partial \,\varrho_\varphi }{\partial\,\varphi} = \frac{1}{2}\left[\varrho_\varphi L_\varphi + L_\varphi \varrho_\varphi \right]. \label{SLD}
\end{equation}
Indeed, quantum Fisher information is a central element of quantum metrology, it can be used to extract information on certain parameters that cannot be measured directly \cite{Hu2018}, this quantity has been proposed specifically for evaluating the sensitivity of quantum systems to phase changes \cite{Yu2018,Yi2012,Liu2016}. It is important to note that this value can be obtained directly when the computation of the SLD operator is performed explicitly. In this study, for any state of the form X (e.g., \eqref{matrix rho}), we focus on the explicit expression of the quantum Fisher information. The analytical expression of the SLD in terms of the density matrix of the studied system is obtained by solving the Eq \eqref{SLD}
\begin{equation}
    L_\varphi = 2 \, \int_0^{+\infty} \, e^{-\varrho_\varphi s}\,(\partial_\varphi \varrho_\varphi)  \, e^{-\varrho_\varphi s} ds.
\end{equation}
Based on the decomposition $\varrho_\varphi^m=\varrho_\varphi^{(1)m} \oplus \varrho_\varphi^{(2)m} \oplus \varrho_\varphi^{(3)m} \oplus \varrho_\varphi^{(4)m}$ (with $m \leq 1$) \\ and $e^{-\varrho_\varphi s}= \sum_{m=0}^{\infty}\, \frac{(-s)^m}{m!} \varrho_\varphi^m = e^{-\varrho_\varphi^{(1)} s}\oplus e^{-\varrho_\varphi^{(2)}s} \oplus e^{-\varrho_\varphi^{(3)} s}\oplus e^{-\varrho_\varphi^{(4)}s}$, the expression for the $L_\varphi$ operator is written as follows
\begin{equation}
    L_\varphi =  \mathcal{L}_\varphi^{(1)} +  \mathcal{L}_\varphi^{(2)}+\mathcal{L}_\varphi^{(3)} +  \mathcal{L}_\varphi^{(4)},
\end{equation}
where

\begin{align}
     &\mathcal{L}^{(1)}_\varphi= p^1_0 \eta_0^1 + \sum_{i=1}^3 p^1_i \eta_i^1, \hspace{2cm}\mathcal{L}^{(2)}_\varphi= p^2_0 \eta_0^2 + \sum_{i=1}^3 p^2_i \eta_i^2,  \nonumber \\ &\mathcal{L}^{(3)}_\varphi= p^3_0 \eta_0^3 + \sum_{i=1}^3 p^3_i \eta_i^3,   \hspace{1cm} \text{and} \hspace{1cm} \mathcal{L}^{(4)}_\varphi= p^4_0 \eta_0^4 + \sum_{i=1}^3 p^4_i \eta_i^4, \label{SLD 1234}
\end{align}

which implies that the general expression of the quantum Fisher information can be given as $\mathbf{\textit{F}}(\varrho_\varphi)=\sum_{i=1}^{4}\mathbf{\textit{F}}\left(\varrho_\varphi^{(i)} \right)$, \text{where} $\mathbf{\textit{F}}\left(\varrho_\varphi^{(i)} \right)$ represents the quantum Fisher information related with $\varrho_\varphi^{(i)}$,$i=1, 2, 3, 4$ which is expressed as 
 \begin{align}
    &\mathbf{\textit{F}}\left(\varrho_\varphi^{(1)} \right) = Tr\left((\partial_\varphi \,\varrho_\varphi^{(1)})\mathcal{L}^{(1)}_\varphi\right), \hspace{2cm} \mathbf{\textit{F}}\left(\varrho_\varphi^{(2)} \right) = Tr\left((\partial_\varphi \,\varrho_\varphi^{(2)})\mathcal{L}^{(2)}_\varphi\right), \\
    &\mathbf{\textit{F}}\left(\varrho_\varphi^{(3)} \right) = Tr\left((\partial_\varphi \,\varrho_\varphi^{(3)})\mathcal{L}^{(3)}_\varphi\right), \hspace{2cm} \mathbf{\textit{F}}\left(\varrho_\varphi^{(4)} \right) = Tr\left((\partial_\varphi \,\varrho_\varphi^{(4)})\mathcal{L}^{(4)}_\varphi\right).
\end{align}
Considering the development of the SLD operator $\mathcal{L}^{(1)}_\varphi$ (Eq. \eqref{SLD 1234}) for the state $\varrho_\varphi^{(1)}$ and (Eq. \eqref{rho in fbch rep}), we can verify that
\begin{equation}
    \mathbf{\textit{F}}\left(\varrho_\varphi^{(1)} \right) = p^1_0 (\partial_\varphi \omega^1_0) + \sum_{i=1}^3 (p^1_i \partial_\varphi \omega^1_i), \label{F1}
\end{equation}
$p^1_0$ and $p^1_i$ are uniquely determined by the following relation
\begin{equation}
    \frac{\partial \varrho_\varphi^{(1)}}{\partial \varphi}=\frac{1}{2}\left( \varrho_\varphi^{(1)} \mathcal{L}^{(1)}_\varphi+ \mathcal{L}^{(1)}_\varphi \varrho_\varphi^{(1)} \right).
\end{equation}
A straightforward substitution of the density matrix $\varrho_\varphi^{(1)}$ (Eq. \eqref{rho in fbch rep}) into the operator $\mathcal{L}^{(1)}_\varphi$ expression (Eq. \eqref{SLD 1234}) allows us to show that
\begin{equation}
    \partial_\varphi \omega_0^1 = \omega_0^1 p_0^1+\sum_{i=1}^3 \omega_i^1 p_i^1 \hspace{1cm} \text{and} \hspace{1cm} \partial_\varphi \omega_i^1 =  p_0^1\omega_i^1 +\omega_0^1 p_i^1, \label{drv eta}
\end{equation}
with $i=0, 1, 2, 3$. 

\begin{equation}
    p_0 = \frac{\omega_0^1 \partial_\varphi \omega_0^1 - \sum_{i=1}^3 \omega_i^1 \partial_\varphi \omega_i^1}{(\omega_0^1)^2-\sum_{i=1}^3 (\omega_i^1)^2},  \hspace{1cm} \text{and} \hspace{1cm} p_i=\frac{1}{\omega_0^1} \Bigg[\frac{g^{\alpha\beta} \omega_\alpha^1\left(\omega_\beta^1 \partial_\varphi \omega_i^1-\omega_i^1 \partial_\varphi\omega_\beta^1  \right)}{g^{\alpha\beta} \omega_\alpha^1 \omega_\beta^1} \bigg], \label{p0 pi}
\end{equation}
where $g^{\alpha\beta}$ represents the Minkowski space-time metric, given by the diag(1, -1, -1, -1).Upon substituting Eqs. \eqref{drv eta} and \eqref{p0 pi}  into
Eq \eqref{F1}, we can express $\mathbf{\textit{F}}\left(\varrho_\varphi^{(1)} \right) $ as
\begin{equation}
    \mathbf{\textit{F}}\left(\varrho_\varphi^{(1)} \right) = \frac{(\partial_\varphi \, \omega_0^1)^2}{\omega_0^1} +\frac{1}{\omega_0^1}\left[\frac{\left(g^{\alpha\beta} \omega_\alpha^1\,\partial_\varphi\,\omega_\beta^1\right)^2}{g^{\alpha\beta} \omega_\alpha^1\,\omega_\beta^1}-g^{\alpha\beta} \,(\partial_\varphi\,\omega_\alpha^1)\,(\partial_\varphi\,\omega_\beta^1)\right]. \label{F generale}
\end{equation}
The Eq \eqref{F generale} is valid for mixed states, $(\text{where}\, \omega_0^1 \neq 0 \, \text{and} \, g^{\alpha\beta}\omega_\alpha^1\omega_\beta^1\neq 0 )$. In the case of pure states, where $\varrho_\varphi^{2}=\varrho_\varphi$, the SLD operator can be computed directly as
\begin{equation}
    \mathcal{L}^{(1)}_\varphi = 2\, \partial_\varphi \, \varrho_\varphi^{(1)}. \label{sld pur}
\end{equation}
In particular, the simplified expression for $\mathbf{\textit{F}}\left(\varrho_\varphi^{(1)} \right)$, which becomes
\begin{equation}
    \mathbf{\textit{F}}\left(\varrho_\varphi^{(1)} \right) = (\omega_0^1)^2 + \sum_{i=1}^3 (\omega_i^1)^2.
\end{equation}
Similarly, the quantum Fisher information $\mathbf{\textit{F}}\left(\varrho_\varphi^{(2)} \right)$, $\mathbf{\textit{F}}\left(\varrho_\varphi^{(3)} \right)$ and $\mathbf{\textit{F}}\left(\varrho_\varphi^{(4)} \right)$ associated with density matrices $\varrho_\varphi^{(2)}$, $\varrho_\varphi^{(3)}$ and $\varrho_\varphi^{(4)}$, can be determined in a similar manner, they are expressed as follows
\begin{align}
     &  \mathbf{\textit{F}}\left(\varrho_\varphi^{(2)} \right) = \frac{(\partial_\varphi \, \omega_0^2)^2}{\omega_0^2} +\frac{1}{\omega_0^2}\left[\frac{\left(g^{\alpha\beta} \omega_\alpha^2\,\partial_\varphi\,\omega_\beta^2\right)^2}{g^{\alpha\beta} \omega_\alpha^2\,\omega_\beta^2}-g^{\alpha\beta} \,(\partial_\varphi\,\omega_\alpha^2)\,(\partial_\varphi\,\omega_\beta^2)\right], \label{F2 generale} \\
     &  \mathbf{\textit{F}}\left(\varrho_\varphi^{(3)} \right) = \frac{(\partial_\varphi \, \omega_0^3)^2}{\omega_0^3} +\frac{1}{\omega_0^3}\left[\frac{\left(g^{\alpha\beta} \omega_\alpha^3\,\partial_\varphi\,\omega_\beta^3\right)^2}{g^{\alpha\beta} \omega_\alpha^3\,\omega_\beta^3}-g^{\alpha\beta} \,(\partial_\varphi\,\omega_\alpha^3)\,(\partial_\varphi\,\omega_\beta^3)\right], \label{F3 generale} \\
     & \mathbf{\textit{F}}\left(\varrho_\varphi^{(4)} \right) = \frac{(\partial_\varphi \, \omega_0^4)^2}{\omega_0^4} +\frac{1}{\omega_0^4}\left[\frac{\left(g^{\alpha\beta} \omega_\alpha^4\,\partial_\varphi\,\omega_\beta^4\right)^2}{g^{\alpha\beta} \omega_\alpha^4\,\omega_\beta^4}-g^{\alpha\beta} \,(\partial_\varphi\,\omega_\alpha^4)\,(\partial_\varphi\,\omega_\beta^4)\right]. \label{F4 generale}
\end{align}
For pure states, the quantities of quantum Fisher information $\mathbf{\textit{F}}\left(\varrho_\varphi^{(2)} \right)$, $ \mathbf{\textit{F}}\left(\varrho_\varphi^{(3)} \right)$ and $ \mathbf{\textit{F}}\left(\varrho_\varphi^{(4)} \right)$ take on the following forms

\begin{align}
    &\mathbf{\textit{F}}\left(\varrho_\varphi^{(2)} \right) = (\omega_0^2)^2 + \sum_{i=1}^3 (\omega_i^2)^2,\\
    &\mathbf{\textit{F}}\left(\varrho_\varphi^{(3)} \right) = (\omega_0^3)^2 + \sum_{i=1}^3 (\omega_i^3)^2, \\
    &\mathbf{\textit{F}}\left(\varrho_\varphi^{(4)} \right) = (\omega_0^4)^2 + \sum_{i=1}^3 (\omega_i^4)^2.  
\end{align}
For pure states, these new expressions \eqref{F2 generale}, \eqref{F3 generale} and \eqref{F4 generale} offer a significant computational advantage over traditional techniques that rely on diagonalizing the density matrix. By avoiding diagonalization, These expressions can be successfully put to use in a wide range of quantum systems.

\subsection{Detailed expression of Skew information}
The skew information stands out as a specific type of quantum Fisher information. This connection arises from its unique property; it behaves as a monotonic metric within the state space \cite{Petz1996, Petz21996}. Additionally, skew information exhibits numerous geometric features mirroring those of Fisher information. This shared foundation in the concept of quantum uncertainty explains their similarity. Furthermore, the Cramér-Rao inequality can be reformulated using skew information \cite{Petz31996,Luo12004}. These observed parallels between skew information and Fisher information fueled our investigation into their deeper connections. In the realm of quantum mechanics, the overall uncertainty attributed to a single observable $\mathrm{K}$ within a quantum state $\varrho_\varphi$ is conventionally measured through the variance $\mathrm{Var}(\varrho_\varphi, \mathrm{K}) = Tr\left(\varrho_\varphi\, \mathrm{K}^2\right) - (Tr(\varrho_\varphi\, \mathrm{K}^2))^2$. However, for mixed states (mixtures of pure states), the variance incorporates two components; classical uncertainty $\mathrm{Var}_c(\varrho_\varphi)$ due to our lack of knowledge about the underlying classical mixture, and a purely quantum mechanical uncertainty $\mathrm{Var}_q(\varrho_\varphi)$. The total variance can be expressed as
\begin{equation}
     \mathrm{Var}(\varrho_\varphi, 
     \mathrm{K}) = \mathrm{Var}_c(\varrho_\varphi) + \mathrm{Var}_q(\varrho_\varphi).
\end{equation}
Wigner and Yanase \cite{WIGNER1963} proposed a measure to isolate the purely quantum component of the variance. The general expression of this measure is given as follows
\begin{equation}
    \mathtt{I}(\varrho_\varphi) = -4 Tr[\sqrt{\varrho_\varphi},\,\mathrm{K} ]^2,\label{I(K)}
\end{equation}
this quantity measures the informational content pertaining to observable values that do not commute with $\mathrm{K}$. In the estimation of an unknown phase shift within the context of quantum estimation \cite{Slaoui2022}, the density matrix is generated by the unitary evolution consistent with the Landau-von Neumann equation $i \frac{\partial}{\partial \varphi} \varrho_\varphi = [\mathrm{K}, \varrho_\varphi]$, and is described by $\varrho_\varphi = e^{-i \mathrm{K} \varphi} \varrho_{ABC} e^{i \mathrm{K} \varphi}$. Therefore, Eq \eqref{I(K)} can be rewritten as
\begin{equation}
    \mathtt{I}(\varrho_\varphi) = 4 Tr\left(\partial\,\sqrt{\varrho_\varphi}\right)^2, \label{I(rho)}
\end{equation}
this metric serves as a gauge for assessing the degree of non-commutativity between an observable $\mathrm{K}$ and the particular state $\varrho_\varphi$. It fulfills all the standard requirements of an information-theoretic measure, demonstrating variance-like behavior for pure states and becoming zero when the conserved observable $\mathrm{K}$ commutes with the density matrix. It remains constant under state transformations for isolated systems, in accordance with the Landau-von Neumann equation. Moreover, the information provided by a general system and the sum of the information provided by subsystems are independent, which means that the skew information is additive. In the case of a three-qubit system \eqref{rho oplus}, the skew information provided can be expressed as follows
\begin{equation}
     \mathtt{I}(\varrho_\varphi) =  \mathtt{I}(\varrho_\varphi^{(1)}) +  \mathtt{I}(\varrho_\varphi^{(2)})+ \mathtt{I}(\varrho_\varphi^{(3)})+ \mathtt{I}(\varrho_\varphi^{(4)}), \label{I(I1 I2 I3 I4)}
\end{equation}
where 
\begin{equation}
\begin{array}{cc}
     & \mathtt{I}\left(\varrho_\varphi^{(1)}\right)= 4 Tr\left(\partial\,\sqrt{\varrho_\varphi^{(1)}}\right)^2, \hspace{2cm} 
     \mathtt{I}\left(\varrho_\varphi^{(2)}\right)= 4 Tr\left(\partial\,\sqrt{\varrho_\varphi^{(2)}}\right)^2,   \\
     & \mathtt{I}\left(\varrho_\varphi^{(3)}\right)= 4 Tr\left(\partial\,\sqrt{\varrho_\varphi^{(3)}}\right)^2, \hspace{2cm} 
     \mathtt{I}\left(\varrho_\varphi^{(4)}\right)= 4 Tr\left(\partial\,\sqrt{\varrho_\varphi^{(4)}}\right)^2. \label{I(rho1234)}
\end{array}
\end{equation}
To derive the explicit expression of $\mathtt{I}\left(\varrho_\varphi^{(1)}\right)$, we use a procedure similar to that used to obtain the quantum Fisher information, with the matrix $\sqrt{\varrho_\varphi^{(1)}}$ being written as follows
\begin{equation}
    \sqrt{\varrho_\varphi^{(1)}} = d_0^1 \eta_0^1+\sum_{i=1}^3 d_i^1 \eta_i^1. \label{sqrt rho}
\end{equation}
Based on the above equation, and utilizing definition \eqref{I(rho1234)}, one can demonstrate that
\begin{equation}
    \mathtt{I}\left(\varrho_\varphi^{(1)}\right) = 8 \left[\left(\partial_\varphi\,d_0^1 \right)^2+\sum_{i=1}^3 \left(\partial_\varphi\,d_i^1 \right)^2 \right]. \label{I d} 
\end{equation}

So, it becomes evident that to compute $\mathtt{I}\left(\varrho_\varphi^{(1)}\right)$, we must determine the coefficients $d_0^1$ and $d_i^1$ in terms of $\omega_0^1$  and $\omega_i^1$ \eqref{ommega 1,2}, respectively. Regarding the matrix $\sqrt{\varrho_\varphi^{(1)}}$, we obtain
\begin{equation}
    \varrho_\varphi^{(1)} = \left(d_0^{1^2}+\left|d^1\right|^2 \right) \eta_0^1+2 \, d_0^1 \sum_{i=1}^3\,d_i^1 \, \eta_i^1.
\end{equation}
We now compare the preceding expression with the explicit formulation of matrix $\varrho_{ABC}^{(1)}$  \eqref{rho in fbch rep}, we find
\begin{equation}
    d_0^{1^2}+\left|d^1\right|^2=\frac{\omega_0^1}{2}\hspace{0.8cm} \,\text{and}\, \hspace{0.8cm} 2 d_0^{1} d_i^{1}= \frac{\omega_i^1}{2},
\end{equation}
with solutions $d_0^1$ and $d_i^1$ given by
\begin{equation}
    d_0^1 = \frac{1}{2} \sqrt{\omega_0^1+\sqrt{\omega_0^{1^2}-\left|\omega^1\right|^2}}, \hspace{2cm} d_i^1 = \frac{1}{2} \frac{\omega_i^1}{\sqrt{\omega_0^1+\sqrt{\omega_0^{1^2}-\left|\omega^1\right|^2}}},\label{d0 di}
\end{equation}
where $\left|\omega^1\right|^2 = \sum_{i=1}^3 \omega_i^{1^2}$ and $\left|d^1\right|^2 = \sum_{i=1}^3 d_i^{1^2}$,
We substitute solution \eqref{d0 di} into Eq \eqref{sqrt rho}, resulting in the matrix $\sqrt{\varrho_\varphi^{(1)}}$ taking the following form
\begin{equation}
    \sqrt{\varrho_\varphi^{(1)}} = \frac{1}{2}\left[\frac{1}{2} \sqrt{\omega_0^1+\sqrt{\omega_0^{1^2}-\left|\omega^1\right|^2}} \, \eta_0^1+\sum_{i=1}^3 \frac{1}{2} \frac{\omega_i^1}{\sqrt{\omega_0^1+\sqrt{\omega_0^{1^2}-\left|\omega^1\right|^2}}}  \, \eta_i^1 \right].
\end{equation}
To derive $d_0^1$ and $d_i^1$, we utilize the expressions from Eq \eqref{d0 di}. Therefore, the expressions for $\partial\varphi\, d_0^1$ and $\partial\varphi\, d_i^1$ are as follows
\begin{equation}
    \partial\varphi\, d_0^1 = \frac{1}{4\,\sqrt{\omega_0^{1^2}-\left|\omega^1\right|^2}} \left[ \sqrt{\omega_0^1+\sqrt{\omega_0^{1^2}-\left|\omega^1\right|^2}} \,\left(\partial\varphi\,\omega_0^1\right) - \frac{\sum_{j=1}^3 \omega_j^1 \left(\partial\varphi\,\omega_j^1\right)}{\sqrt{\omega_0^1+\sqrt{\omega_0^{1^2}-\left|\omega^1\right|^2}}} \right], \label{partial d0}
\end{equation}
and
\begin{equation}
     \partial\varphi\, d_i^1 = \frac{-\sigma}{4} \left(\omega_i^1 \partial\varphi\,\omega_0^1 \right) + \frac{\lambda}{2} \left( \partial\varphi\,\omega_i^1 \right) + \frac{\Gamma}{4} \omega_i^1 \sum_{j=1}^3 \left( \omega_j^1 \partial\varphi\,\omega_j^1\right), \label{partial di}
\end{equation}
where $\sigma$, $\lambda$ and $\Gamma$ are defined as
\begin{align}
    \sigma =& \left(\omega_0^{1^2} - \left|\omega^1\right|^2 \right)^{-\frac{1}{2}} \left(\omega_0^1 + \sqrt{\omega_0^{1^2} - \left|\omega^1\right|^2} \right)^{-\frac{1}{2}},  \nonumber\\ 
    \lambda =& \left(\omega_0^1 +\sqrt{\omega_0^{1^2} - \left|\omega^1\right|^2} \right)^{-\frac{1}{2}}, \nonumber \\
    \Gamma =& \left(\omega_0^{1^2} - \left|\omega^1\right|^2 \right)^{-\frac{1}{2}} \left(\omega_0^1 + \sqrt{\omega_0^{1^2} - \left|\omega^1\right|^2} \right)^{-\frac{3}{2}}. \nonumber
\end{align}

The substitution of Eqs  \eqref{partial d0} and \eqref{partial di} into Eq \eqref{I d} provides the necessary expression for the skew information $\mathtt{I}\left(\varrho_\varphi^{(1)}\right)$. To obtain the expressions for $\mathtt{I}\left(\varrho_\varphi^{(2)}\right)$, $\mathtt{I}\left(\varrho_\varphi^{(3)}\right)$ and $\mathtt{I}\left(\varrho_\varphi^{(4)}\right)$, we replace the coefficient $\omega^1_{\alpha}$ with $\omega^2_{\alpha}$, $\omega^3_{\alpha}$ and $\omega^4_{\alpha}$. The sum of these expressions gives the skew information $\mathtt{I}\left(\varrho_\varphi\right)$ (Eq. \eqref{I(I1 I2 I3 I4)}).

\section{Deriving explicit formulas for quantum Fisher and Skew information  under decoherence effect}
Environmental decoherence describes the unwanted interaction between a qubit and its surroundings. These environmental effects can significantly impact the efficiency of manipulating quantum information 
 \cite{Hornberger2009,Maziero2009}. To reflect our understanding of the environmental influences, it is crucial to have an accurate mathematical description of the errors\cite{Bourennane2004}. In noisy environments, how quantum information changes over time can be described using mathematical tools called quantum channels. In fact, a quantum channel denoted by $\phi$, is a mathematical function that transforms a pure state into a mixed state due to the influence of its surroundings \cite{Hornberger2009,Schlosshauer2007}. In general, a quantum channel is fully described by the Kraus representation as follows
 \begin{equation}
     \phi (\varrho) = \sum_{\mu} \mathtt{K}_\mu \,\varrho\, \mathtt{K}^\dagger_\mu,
 \end{equation}
 where $\mathtt{K}_\mu$ are the Kraus operators that meet the following conditions
 \begin{equation}
     \sum_{\mu} \mathtt{K}^\dagger_\mu \,\mathtt{K}_\mu = \mathbb{I},
 \end{equation}
 where $\mathbb{I}$ is a matrix identity. In the Fano-Bloch representation, the effect of decoherence on a density matrix $\varrho$ representing a three-qubit system may be expressed as follows
 \begin{equation}
     \phi (\varrho) = \sum_{\alpha\beta\gamma} \, \Sigma (\mathcal{T})_{\alpha \beta \gamma} \,\sigma_\alpha \otimes \sigma_\beta \otimes \sigma_\gamma,
 \end{equation}
 where the Pauli matrices with $i = 1, 2, 3$ are denoted by $\sigma_i$, and elements of the correlation matrix are described by 
 \begin{equation}
    \phi\left( (\mathcal{T})_{\alpha \beta \gamma}\right) = Tr\left( \phi^\dagger (\phi_\alpha) \otimes \phi^\dagger (\phi_\beta) \otimes \phi^\dagger (\phi_\gamma) \right), \label{phi(T)}
 \end{equation}
 with
 \begin{equation}
     \phi^\dagger (\phi_\alpha) = \sum_{\mu} \mathtt{K}^\dagger_\mu \phi_\alpha \mathtt{K}_\mu.
 \end{equation}
 We now study the skew information and quantum Fisher information for a three-qubit system in the three decoherence channels (phase damping, depolarization, and phase flip). This system may be represented by Kraus operators such as the ones below
 
 \textbf{a. Phase damping channel;} this channel is characterized by the following Kraus operators
 \begin{equation}
     \mathtt{K}^{PDC}_0 = \sqrt{S} \, \mathbb{I}, \hspace{1cm} \mathtt{K}^{PDC}_1 = \sqrt{p} \, |0\rangle\langle0|, \hspace{1cm} \mathtt{K}^{PDC}_2 = \sqrt{p} \, |1\rangle\langle1|, \label{K(PDC)}
 \end{equation}
where the probability of decoherence $p$ is determined by  $p=1-S$, with $S=1-e^{-\gamma t}$, where $\gamma$ is the decay rate. This channel presents a prototype model for pure decoherence, which is defined as the degradation of coherence in a three-level system while conserving the system energy.

\textbf{b. Depolarizing channel;} 
this channel represents a decoherence model in which the qubit becomes completely mixed with a probability of $3p/4$, The following set provides the four Kraus operators defining this process
\begin{equation}
     \mathtt{K}^{DPC}_0 = \sqrt{1-p'} \, \mathbb{I}, \hspace{1cm} \mathtt{K}^{DPC}_1 = \sqrt{\frac{p'}{3}} \, \sigma_x , \hspace{1cm} \mathtt{K}^{DPC}_2 = \sqrt{\frac{p'}{3}} \, \sigma_y , \hspace{1cm} \mathtt{K}^{DPC}_3 = \sqrt{\frac{p'}{3}} \, \sigma_z ,\label{K(DPC)}
\end{equation}
with $p'=3p/4$.

\textbf{c. Phase flip channel;} 
quantum noise manifests in various forms, one of which is the phase flip channel that can randomly flip the sign of a qubit state. Imagine it like a switch that can reverse the orientation of the information the qubit holds, the corresponding Kraus operators are given by

\begin{equation}
    \mathtt{K}^{PFC}_0 = \sqrt{1-p} \, \mathbb{I}, \hspace{2cm} \mathtt{K}^{PFC}_1 = \sqrt{p} \, \sigma_z .\label{K(PFC)}
\end{equation}
The three channels described above encompass a wide range of decoherence types encountered in experiments. The quantum Fisher information and skew information associated with the Greenberger–Horne–Zeilinger (GHZ) state within these channels are analyzed in the following sections.
\subsection{Quantum Fisher and Skew information for three qubit X-states}
Starting from the Kraus operators, which define the three quantum channels explained earlier, we will discuss in this part of the work the analytical expressions of quantum Fisher information and skew information in the three quantum channels for three-qubit X-states. It is clear that the X-form of the density matrix cannot be modified by decoherence. To show that the decomposition of a density matrix into four submatrices under the effect of decoherence is possible, to verify it, we use the linearity property $\phi(\alpha \varrho_1 + \beta \varrho_2 + \gamma \varrho_3 + \sigma \varrho_4) = \alpha \phi(\varrho_1) + \beta \phi(\varrho_2) + \gamma \phi(\varrho_3) + \sigma \phi(\varrho_4)$, and the expression from Eq. \eqref{rho oplus}. We observe that
\begin{equation}
    \phi(\varrho_{ABC}) =  \phi(\varrho_{ABC}^{(1)}) + \phi(\varrho_{ABC}^{(2)}) + \phi(\varrho_{ABC}^{(3)}) + \phi(\varrho_{ABC}^{(4)}), \label{Sigma 1234}
\end{equation}
where $\phi(\varrho_{ABC}^{(1)})$, $\phi(\varrho_{ABC}^{(2)})$, $\phi(\varrho_{ABC}^{(3)})$ and $\phi(\varrho_{ABC}^{(4)})$ are given by
\begin{align}
    &\phi(\varrho_{ABC}^{(1)}) = \frac{1}{2}\, \sum_{\alpha=0}^{3}\Omega_\alpha^1 \eta_\alpha^1, \hspace{1cm} \phi(\varrho_{ABC}^{(2)}) = \frac{1}{2}\,  \sum_{\alpha=0}^{3}\Omega_\alpha^2 \eta_\alpha^2,\\
   &\phi(\varrho_{ABC}^{(3)}) = \frac{1}{2}\,  \sum_{\alpha=0}^{3}\Omega_\alpha^3 \eta_\alpha^3, \hspace{1cm} \phi(\varrho_{ABC}^{(4)}) = \frac{1}{2}\,  \sum_{\alpha=0}^{3}\Omega_\alpha^4 \eta_\alpha^4.
\end{align}
Thus, the quantum Fisher information corresponding to the total state of the system under the effect of decoherence is written as follows
\begin{equation}
\mathbf{\textit{F}}\big(\phi\left(\varrho_{ABC}\right)\big) = \mathbf{\textit{F}}\left(\phi\left(\varrho_{ABC}^{(1)}\right) \right)+\mathbf{\textit{F}}\left(\phi\left(\varrho_{ABC}^{(2)}\right) \right)+\mathbf{\textit{F}}\left(\phi\left(\varrho_{ABC}^{(3)}\right) \right)+\mathbf{\textit{F}}\left(\phi\left(\varrho_{ABC}^{(4)}\right) \right),
\end{equation}
where
\begin{align}
&  \mathbf{\textit{F}}\left(\phi(\varrho_{ABC}^{(1)}) \right) = \frac{(\partial_\varphi \, \Omega_0^1)^2}{\Omega_0^1} +\frac{1}{\Omega_0^1}\left[\frac{\left(g^{\alpha\beta} \Omega_\alpha^1\,\partial_\varphi\,\Omega_\beta^1\right)^2}{g^{\alpha\beta} \Omega_\alpha^1\,\Omega_\beta^1}-g^{\alpha\beta} \,(\partial_\varphi\,\Omega_\alpha^1)\,(\partial_\varphi\,\Omega_\beta^1)\right], \label{F1 rho} \\
     &  \mathbf{\textit{F}}\left(\phi(\varrho_{ABC}^{(2)}) \right) = \frac{(\partial_\varphi \, \Omega_0^2)^2}{\Omega_0^2} +\frac{1}{\Omega_0^2}\left[\frac{\left(g^{\alpha\beta} \Omega_\alpha^2\,\partial_\varphi\,\Omega_\beta^2\right)^2}{g^{\alpha\beta} \Omega_\alpha^2\,\Omega_\beta^2}-g^{\alpha\beta} \,(\partial_\varphi\,\Omega_\alpha^2)\,(\partial_\varphi\,\Omega_\beta^2)\right], \label{F2 rho} \\
     &  \mathbf{\textit{F}}\left(\phi(\varrho_{ABC}^{(3))} \right) = \frac{(\partial_\varphi \, \Omega_0^3)^2}{\Omega_0^3} +\frac{1}{\Omega_0^3}\left[\frac{\left(g^{\alpha\beta} \Omega_\alpha^3\,\partial_\varphi\,\Omega_\beta^3\right)^2}{g^{\alpha\beta} \Omega_\alpha^3\,\Omega_\beta^3}-g^{\alpha\beta} \,(\partial_\varphi\,\Omega_\alpha^3)\,(\partial_\varphi\,\Omega_\beta^3)\right], \label{F3 rho} \\
     & \mathbf{\textit{F}}\left(\phi(\varrho_{ABC}^{(4)}) \right) = \frac{(\partial_\varphi \, \Omega_0^4)^2}{\Omega_0^4} +\frac{1}{\Omega_0^4}\left[\frac{\left(g^{\alpha\beta} \Omega_\alpha^4\,\partial_\varphi\,\Omega_\beta^4\right)^2}{g^{\alpha\beta} \Omega_\alpha^4\,\Omega_\beta^4}-g^{\alpha\beta} \,(\partial_\varphi\,\Omega_\alpha^4)\,(\partial_\varphi\,\Omega_\beta^4)\right]. \label{F4 rho}
\end{align}
 For pure states, Eqs \eqref{F1 rho}, \eqref{F2 rho}, \eqref{F3 rho} and \eqref{F4 rho} take the following simplified form
 \begin{align}
&\mathbf{\textit{F}}\left(\phi(\varrho_{ABC}^{(1)}) \right) = (\Omega_0^1)^2 + \sum_{i=1}^3 (\Omega_i^1)^2, \hspace{1cm}
 \mathbf{\textit{F}}\left(\phi(\varrho_{ABC}^{(2)}) \right) = (\Omega_0^2)^2 + \sum_{i=1}^3 (\Omega_i^2)^2,\\
    &\mathbf{\textit{F}}\left(\phi(\varrho_{ABC}^{(3)}) \right) = (\Omega_0^3)^2 + \sum_{i=1}^3 (\Omega_i^3)^2,\hspace{1cm}
   \mathbf{\textit{F}}\left(\phi(\varrho_{ABC}^{(4)}) \right) = (\Omega_0^4)^2 + \sum_{i=1}^3 (\Omega_i^4)^2. 
\end{align}
We know that the form X of the three-qubit density matrix remains unchanged under the effect of decoherence. Therefore, states $\sqrt{\phi(\varrho_{ABC}^{(j)})}$ (with $j=1,...,4$) are expressed as a combination of generators $\eta_\alpha^1$, $\eta_\alpha^2$,$\eta_\alpha^3$ and $\eta_\alpha^4$ as follows
\begin{equation}
    \sqrt{\phi(\varrho_{ABC}^{(1)})} = \sum_{\alpha=1}^3 X_\alpha^1\, \eta_\alpha^1, \hspace{0.2cm} \sqrt{\phi(\varrho_{ABC}^{(2)})} = \sum_{\alpha=1}^3 X_\alpha^2\, \eta_\alpha^2, \hspace{0.2cm} \sqrt{\phi(\varrho_{ABC}^{(3)})} = \sum_{\alpha=1}^3 X_\alpha^3\, \eta_\alpha^3 
,\hspace{0.2cm}\sqrt{\phi(\varrho_{ABC}^{(4)})} = \sum_{\alpha=1}^3 X_\alpha^4\, \eta_\alpha^4, \label{sqrt phi rho}
\end{equation}

where $X_\alpha^1$, $X_\alpha^2$, $X_\alpha^3$ and $X_\alpha^4$ are the elements of the Bloch vector for the density matrices 
$\sqrt{\phi(\varrho_{ABC}^{(1)})}$, $\sqrt{\phi(\varrho_{ABC}^{(2)})}$, $\sqrt{\phi(\varrho_{ABC}^{(3)})}$ and $\sqrt{\phi(\varrho_{ABC}^{(4)})}$ respectively. Consequently, the skew information expressions corresponding to these matrices are defined as follows

\begin{align}
    \mathtt{I}\left(\phi(\varrho_{ABC}^{(1)})\right) =& 8 \left[\left(\partial_\varphi\,X_0^1 \right)^2+\sum_{j=1}^3 \left(\partial_\varphi\,X_j^1 \right)^2 \right], \hspace{1.5cm}  \mathtt{I}\left(\phi(\varrho_{ABC}^{(2)})\right) = 8 \left[\left(\partial_\varphi\,X_0^2 \right)^2+\sum_{j=1}^3 \left(\partial_\varphi\,X_j^2 \right)^2 \right], \\
    \mathtt{I}\left(\phi(\varrho_{ABC}^{(3)})\right) =& 8 \left[\left(\partial_\varphi\,X_0^3 \right)^2+\sum_{j=1}^3 \left(\partial_\varphi\,X_j^3 \right)^2 \right], \hspace{1.5cm}  \mathtt{I}\left(\phi(\varrho_{ABC}^{(4)})\right) = 8 \left[\left(\partial_\varphi\,X_0^4 \right)^2+\sum_{j=1}^3 \left(\partial_\varphi\,X_j^4 \right)^2 \right].
\end{align}

To obtain the detailed expressions of the quantum Fisher information and skew information for any three-qubit system, we can apply the operator $\phi$ to the matrix $\mathcal{T}$, this yields $\phi(\mathcal{T})$, which represents a correlation matrix under decoherence. Then, the general form of this matrix in the \textbf{phase-damping channel} is given as follows

\begin{equation}
		\phi(\mathcal{T})^{PDC}=\left(\begin{array}{cccccccc}
               \mathcal{T}_{000} & 0 & 0 & 0 & 0 & 0 & 0 & S^3\mathcal{T}_{111}  \\
			0 & \mathcal{T}_{003} & 0 & 0 & 0 & 0 & S^3\mathcal{T}_{112} & 0 \\
			0 & 0 & \mathcal{T}_{030} & 0 & 0 & S^3\mathcal{T}_{121} & 0 & 0 \\
			0 & 0 & 0 & \mathcal{T}_{033} & S^3\mathcal{T}_{122} & 0 & 0 & 0 \\
                0 & 0 & 0 & S^3\mathcal{T}_{211} & \mathcal{T}_{300} & 0 & 0 & 0 \\
                0 & 0 & S^3\mathcal{T}_{212} & 0 & 0 & \mathcal{T}_{303} & 0 & 0 \\
                0 & S^3\mathcal{T}_{221} & 0 & 0 & 0 & 0 & \mathcal{T}_{330} & 0 \\
                S^3\mathcal{T}_{222} & 0 & 0 & 0 & 0 & 0 & 0 & \mathcal{T}_{333} \\
   
		\end{array}\right), \label{matrix T(PDC)}
				\end{equation}
and the quantities 
$\Omega^1_i, \Omega^2_i, \Omega^3_i \, \text{and} \, \Omega^4_i$ are provided as follows

\begin{align}
\left\{
\begin{array}{ll}
          \Omega^1_0=\omega_0^1,\\
          \Omega^1_1= S^3\omega_1^1, \\
          \Omega^1_2 = S^3\omega_2^1, \\
          \Omega^1_3 = \omega_3^1,  
\end{array} 
\right.
\hspace{1.5cm}
\left\{
\begin{array}{ll}
          \Omega^2_0=\omega_0^2,\\
          \Omega^2_1= S^3\omega_1^2, \\
          \Omega^2_2 = S^3\omega_2^2, \\
          \Omega^2_3 = \omega_3^2,  
\end{array}  
\right. 
\hspace{1.5cm}
\left\{
\begin{array}{ll}
          \Omega^3_0=\omega_0^3,\\
          \Omega^3_1= S^3\omega_1^3, \\
          \Omega^3_2 = S^3\omega_2^3, \\
          \Omega^3_3 = \omega_3^3,  
\end{array} 
\right.
\hspace{1.5cm}
\left\{
\begin{array}{ll}
          \Omega^4_0=\omega_0^4,\\
          \Omega^4_1= S^3\omega_1^4, \\
          \Omega^4_2 = S^3\omega_2^4, \\
          \Omega^4_3 = \omega_3^4.  
\end{array} 
\right.
\end{align}

Similarly, it is easy to apply $\phi$ to the $\mathcal{T}$ matrix for the \textbf{depolarizing channel}, and we find that 

\begin{equation}
		\phi(\mathcal{T})^{ADC}=\left(\begin{array}{cccccccc}
               \mathcal{T}_{000} & 0 & 0 & 0 & 0 & 0 & 0 & S^3\mathcal{T}_{111}  \\
			0 & S\mathcal{T}_{003} & 0 & 0 & 0 & 0 & S^3\mathcal{T}_{112} & 0 \\
			0 & 0 & S\mathcal{T}_{030} & 0 & 0 & S^3\mathcal{T}_{121} & 0 & 0 \\
			0 & 0 & 0 & S^2\mathcal{T}_{033} & S^3\mathcal{T}_{122} & 0 & 0 & 0 \\
                0 & 0 & 0 & S^3\mathcal{T}_{211} & S\mathcal{T}_{300} & 0 & 0 & 0 \\
                0 & 0 & S^3\mathcal{T}_{212} & 0 & 0 & S^2\mathcal{T}_{303} & 0 & 0 \\
                0 & S^3\mathcal{T}_{221} & 0 & 0 & 0 & 0 & S^2\mathcal{T}_{330} & 0 \\
                S^3\mathcal{T}_{222} & 0 & 0 & 0 & 0 & 0 & 0 & S^3\mathcal{T}_{333} \\
		\end{array}\right), \label{matrix T(DPC)}
				\end{equation}
    and in this channel we obtain

\begin{align}
&\left\{
\begin{array}{ll}
          \Omega^1_0=\frac{1}{4}((1+3S^2)\omega_0^1 + (1-S^2)(\omega_0^2+\omega_0^3+\omega_0^4)),\\
          \Omega^1_1= S^3\omega_1^1, \\
          \Omega^1_2 = S^3\omega_2^1, \\
          \Omega^1_3 = S\omega_3^1,  
\end{array} 
\right.
\left\{
\begin{array}{ll}
          \Omega^2_0=\frac{1}{4}((1+3S^2)\omega_0^2 + (1-S^2)(\omega_0^1+\omega_0^3+\omega_0^4)),\\
          \Omega^2_1= S^3\omega_1^2, \\
          \Omega^2_2 = S^3\omega_2^2, \\
          \Omega^2_3 = S\omega_3^2,  
\end{array}  
\right. \\
\text{and} \nonumber \\
&\left\{
\begin{array}{ll}
          \Omega^3_0=\frac{1}{4}((1+3S^2)\omega_0^3 + (1-S^2)(\omega_0^1+\omega_0^2+\omega_0^4)),\\
          \Omega^3_1= S^3\omega_1^3, \\
          \Omega^3_2 = S^3\omega_2^3, \\
          \Omega^3_3 = S\omega_3^3,  
\end{array} 
\right.
\left\{
\begin{array}{ll}
          \Omega^4_0=\frac{1}{4}((1+3S^2)\omega_0^4 + (1-S^2)(\omega_0^1+\omega_0^2+\omega_0^3)),\\
          \Omega^4_1= S^3\omega_1^4, \\
          \Omega^4_2 = S^3\omega_2^4, \\
          \Omega^4_3 = S\omega_3^4.  
\end{array} 
\right.
\end{align}

For the \textbf{phase flip channel}, the correlation matrix $\phi(\mathcal{T})$ takes the following form
\begin{equation}
		\phi(\mathcal{T})^{PFC}=\left(\begin{array}{cccccccc}
               \mathcal{T}_{000} & 0 & 0 & 0 & 0 & 0 & 0 & A\mathcal{T}_{111}  \\
			0 & \mathcal{T}_{003} & 0 & 0 & 0 & 0 & A\mathcal{T}_{112} & 0 \\
			0 & 0 & \mathcal{T}_{030} & 0 & 0 & A\mathcal{T}_{121} & 0 & 0 \\
			0 & 0 & 0 & \mathcal{T}_{033} & A\mathcal{T}_{122} & 0 & 0 & 0 \\
                0 & 0 & 0 & A\mathcal{T}_{211} & \mathcal{T}_{300} & 0 & 0 & 0 \\
                0 & 0 & A\mathcal{T}_{212} & 0 & 0 & \mathcal{T}_{303} & 0 & 0 \\
                0 & A\mathcal{T}_{221} & 0 & 0 & 0 & 0 & \mathcal{T}_{330} & 0 \\
                A\mathcal{T}_{222} & 0 & 0 & 0 & 0 & 0 & 0 & \mathcal{T}_{333} \\
		\end{array}\right), \label{matrix T(PFC)}
				\end{equation}
where $A=(2S-1)^3$ and 

\begin{align}
\left\{
\begin{array}{ll}
          \Omega^1_0=\omega_0^1,\\
          \Omega^1_1= (2S-1)^3\omega_1^1, \\
          \Omega^1_2 = (2S-1)^3\omega_2^1, \\
          \Omega^1_3 = \omega_3^1,  
\end{array} 
\right.
\hspace{0.5cm}
\left\{
\begin{array}{ll}
          \Omega^2_0=\omega_0^2,\\
          \Omega^2_1= (2S-1)^3\omega_1^2, \\
          \Omega^2_2 = (2S-1)^3\omega_2^2, \\
          \Omega^2_3 = \omega_3^2,  
\end{array}  
\right. 
\hspace{0.5cm}
\left\{
\begin{array}{ll}
          \Omega^3_0=\omega_0^3,\\
          \Omega^3_1= (2S-1)^3\omega_1^3, \\
          \Omega^3_2 = (2S-1)^3\omega_2^3, \\
          \Omega^3_3 = \omega_3^3,  
\end{array} 
\right.
\hspace{0.5cm}
\left\{
\begin{array}{ll}
          \Omega^4_0=\omega_0^4,\\
          \Omega^4_1= (2S-1)^3\omega_1^4, \\
          \Omega^4_2 = (2S-1)^3\omega_2^4, \\
          \Omega^4_3 = \omega_3^4.  
\end{array} 
\right.
\end{align}

As previously mentioned, the correlation matrices $\phi(\mathcal{T})$ cannot be modified in their form X in the three noisy channels (see Eqs \eqref{matrix T(PDC)}, \eqref{matrix T(DPC)}, and \eqref{matrix T(PFC)}). In the next paragraph, we will calculate and explain the evolution of the quantum Fisher and skew information of the GHZ state in the deferent channels.

\section{Application to Werner-GHZ states}
Maximally entangled states of three or more particles, known as Greenberger-Horne-Zeilinger (GHZ) states \cite{Greenberger1990}, have been intriguing quantum systems for exploring the nonlocality inherent in quantum mechanics. Recently, the emerging field of quantum information theory \cite{Bennett1995} has demonstrated that these quantum entangled states can be utilized for tasks such as information transmission and processing. Examples include quantum dense coding \cite{Bennett1992, Mattle1996} and quantum teleportation \cite{Bennett1993, Bouwmeester1997}. The general expression for three-qubit GHZ mixed states is given by
\begin{equation}
    \varrho_{GHZ} = \frac{q}{8} \, \mathbb{I}_3 + (1-q) \left|GHZ\rangle  \langle GHZ\right|, \label{rho(GHZ)}
\end{equation}
where the state $|GHZ\rangle$ is given by
\begin{equation}
    |GHZ\rangle = \frac{1}{\sqrt{2}} \left( \left|000\rangle + \right| 111 \rangle  \right) ,
\end{equation}
In the computational basis $\left\{ |000\rangle, |001\rangle, |010\rangle, |011\rangle, |100\rangle, |101\rangle, |110\rangle, |111\rangle \right\}$, the state given by Eq \eqref{rho(GHZ)} takes the form
\begin{equation}
		\varrho_{GHZ}=\frac{1}{8}\left(\begin{array}{cccccccc}
               4-3\,q & 0 & 0 & 0 & 0 & 0 & 0 & 4(1-q)  \\
			0 & q & 0 & 0 & 0 & 0 & 0 & 0 \\
			0 & 0 & q & 0 & 0 & 0 & 0 & 0 \\
			0 & 0 & 0 & q & 0 & 0 & 0 & 0 \\
                0 & 0 & 0 & 0 & q & 0 & 0 & 0 \\
                0 & 0 & 0 & 0 & 0 & q & 0 & 0 \\
                0 & 0 & 0 & 0 & 0 & 0 & q & 0 \\
                4(1-q) & 0 & 0 & 0 & 0 & 0 & 0 & 4-3\,q \\
		\end{array}\right), \label{rho(GHZ)}
\end{equation}
    with the non-zero elements of the correlation matrix $\mathcal{T}_{\alpha \beta \gamma}$ (with $\alpha, \beta, \gamma = 0, 1, 2, 3$) given as follows
    \begin{equation}
\begin{array}{cc}
    &\mathcal{T}_{000} = 1 , \hspace{0.5cm}  \mathcal{T}_{033} = 1-\text{q}, \hspace{1cm} \mathcal{T}_{303} = 1-\text{q}, \hspace{1cm} \mathcal{T}_{330} = 1-\text{q}, \\
    &\mathcal{T}_{111} = 1-\text{q}, \hspace{0.5cm} \mathcal{T}_{122} = -(1-\text{q}), \hspace{0.5cm} \mathcal{T}_{212} = -(1-\text{q}), \hspace{0.5cm} \mathcal{T}_{221} = -(1-\text{q}).
\end{array}
 \end{equation}
Now, we can decompose this density matrix as follows
\begin{equation}
    \varrho_{GHZ} =  \zeta^1 \oplus \zeta^2 \oplus \zeta^3 \oplus \zeta^4 ,
\end{equation}
where
\begin{equation}
    \zeta^1 = \frac{1}{2} \,\sum_{\alpha=0}^3 \omega^1_\alpha \eta^1_\alpha, \hspace{0.5cm}  \zeta^2 = \frac{1}{2} \,\sum_{\alpha=0}^3 \omega^2_\alpha \eta^2_\alpha, \hspace{0.5cm}  \zeta^3 = \frac{1}{2} \,\sum_{\alpha=0}^3 \omega^3_\alpha \eta^3_\alpha, \hspace{0.2cm} \text{and} \hspace{0.2cm}  \zeta^4 = \frac{1}{2} \,\sum_{\alpha=0}^3 \omega^4_\alpha \eta^4_\alpha. 
\end{equation}

The quantities $\omega^i_\alpha$ (with $i= 1, 2, 3, 4$ and $\alpha = 0, 1, 2, 3$) are given by

\begin{align}
\left\{
\begin{array}{ll}
          \omega^1_0=\frac{4-3 \,q}{4},\\
          \omega^1_1= 1-q, \\
          \omega^1_2 = 0, \\
          \omega^1_3 = 0,  
\end{array} 
\right.
\hspace{1cm}
\left\{
\begin{array}{ll}
          \omega^2_0=\frac{q}{4},\\
          \omega^2_1= 0, \\
          \omega^2_2 = 0, \\
          \omega^2_3 =0,  
\end{array}  
\right. 
\hspace{1cm}
\left\{
\begin{array}{ll}
          \omega^3_0=\frac{q}{4},\\
          \omega^3_1= 0, \\
          \omega^3_2 = 0, \\
          \omega^3_3 =0,  
\end{array} 
\right.
\hspace{1cm}
\left\{
\begin{array}{ll}
          \omega^4_0=\frac{q}{4},\\
          \omega^4_1= 0, \\
          \omega^4_2 = 0, \\
          \omega^4_3 = 0.  
\end{array} 
\right.
\end{align}

\subsection{Dynamics of quantum information metrics in a phase damping channel}
Even after applying this channel, the resulting density matrix $(\phi(\varrho_{GHZ}) =\varrho^{PDC} )$ maintains its X-type nature. This enables us to directly apply the methods from the previous section to assess measures of quantum information such as skew information and quantum Fisher information, to compute the expression of the last quantity for a GHZ state under the effect of a phase-damping channel, we utilize the operators defined in Eq \eqref{K(PDC)} and after some algebraic manipulations, we find that
\begin{equation}
    \mathbf{\textit{F}}\left(\varrho^{PDC} \right) = \frac{9}{4(4-3\,q)}  + \frac{3}{4\,q}+ \frac{4}{4-3\,q} \left[\frac{(-3(4-3\,q)/16+S^6(1-q))^2}{(4-3q)^2/16-S^6(1-q)^2} - \left(\frac{9}{16}-S^6\right)\right].
    \end{equation}
Then, under the phase damping channel, the Wootters concurrence of the GHZ states is calculated as below
\begin{equation}
     \mathbf{C}\left(\varrho^{PDC}\right) = \text{max} \left\{0,\frac{-3\,q}{4} + \frac{1}{8}(4-3q+4S^3-4qS^3)+\frac{1}{8}(-4+3q+4S^3-4qS^3) \right\}.
\end{equation}
Based on the analytical expression of the skew information that we have already demonstrated in the previous section, we find that this criterion for the matrix $\varrho^{PDC}$ is written as follows
\begin{align} \mathtt{I}\left(\varrho^{PDC}\right) =  \frac{3}{4q} + 8 &\Biggr[\left(\frac{1}{4\sqrt{(4-3\,q)^2/16 - S^6(1-q)^2}}\left(\frac{-3\sqrt{k}}{4} + \frac{S^6(1-q)}{\sqrt{k}}\right) \right)^2 \nonumber  \\
 &+  \left(\frac{3\,\sigma_1}{16} S^3 (1-q) - \frac{\lambda_1}{2}\, S^3 - \frac{\Gamma_1}{4}S^9(1-q)^2\right)^2\Biggr], 
\end{align}
with the quantities $k$, $\lambda_1$, $\vartheta_1$\, and $\Gamma_1$ expressed as functions of q (mixing parameter) as follows
\begin{align}
    k=& \frac{1}{4} (4-3\text{q}) + \sqrt{\frac{1}{16}(4-3\,\text{q})^2 - S^6(1-\text{q})^2}, \hspace{0.5cm} \lambda_1 = \left(\frac{1}{4} (4-3\,\text{q}) + \sqrt{\frac{1}{16}(4-3\,\text{q})^2 - S^6(1-\text{q})^2}\right)^{-\frac{1}{2}}, \\
    \vartheta_1 =&\lambda_1 \left(\frac{1}{16}(4-3\,\text{q})^2 - S^6(1-\text{q})^2 \right)^{-\frac{1}{2}} \hspace{0.5cm} \text{and} \hspace{0.25cm} \Gamma_1 = (\lambda_1)^3 \left(\frac{1}{16}(4-3\,\text{q})^2 - S^6(1-\text{q})^2 \right)^{-\frac{1}{2}}.
\end{align}
\subsection{Evolution of quantum criteria under the depolarizing channel}
The analytical expressions of the quantum Fisher information, concurrence entanglement and skew information under the effect of the depolarizing channel are obtained when the operators given by Eq \eqref{K(DPC)} are applied to the GHZ state. It is easy to verify that the resulting density matrix is always of type X. Using the previous results, we find that the expression for the quantum Fisher information subjected to this noise for the GHZ state is expressed as follows
\begin{align}
     \mathbf{\textit{F}}\left(\varrho^{DPC} \right) =& \,  \frac{4}{1+3\,S^2-3\,q\,S^2} \left[\frac{\left(-3\,S^2(1+3\,S^2-3\,q\,S^2)/16 + S^3(S^3-q\,S^3)\right)^2}{(1+3\,S^2-3\,q\,S^2)^2/16-(S^3-q\,S^3)^2}-\left(\frac{9\,S^4}{16}-S^6\right)\right] \nonumber \\
     &+ \frac{3\,S^4}{16\left(1-S^2+q\,S^2 \right)} + \frac{9\,S^4}{4(1+3\,S^2-3\,q\,S^2)}. 
\end{align}
The concurrence entanglement of the GHZ state under the depolarizing effect can be easily expressed by
\begin{align}
     \mathbf{C}\left(\varrho^{DPC}\right) = \text{max} \left\{0, \frac{1}{8}\left(-3\,q\,S(1+S) + \sqrt{16 (1-2\,S^3 + 5\,S^6)-8\,q(3-6S^3+19\,S^6)+q^2(9-18\,S^3+73\,S^6)}  \right)\right\}.
\end{align}
A more compact formula for the skew information of the GHZ state subjected to the depolarizing channel is

\begin{align}
    \mathtt{I}\left(\varrho^{DPC}\right) =& 8 \Bigg[\frac{1}{\left( 1+3\,S^2-3\,q\,S^2 \right)^2 - 16(S^3-q\,S^3)^2}\left(\frac{-3\,S^2}{4} (\lambda_2)^{-1} + \lambda_2 \,S^3(S^3-q\,S^3)\right)^2\nonumber \\
    &+ \left(\frac{3\,\vartheta_2}{16} S^2(S^3-qS^3) - \frac{\lambda_2}{2} S^3 - \frac{\Gamma_2}{4} S^3(S^3-q\,S^3) \right)^2  \Bigg] + \frac{3\,S^4}{4(1-S^2+q\,S^2)}.
\end{align}
With the quantities $\lambda_2$, $ \vartheta_2$ and $\Gamma_2$ defined by
\begin{align}
    \lambda_2 =& \left(\frac{1+S^2}{4}-\frac{3\,q\,S^2}{4}+\sqrt{\left( \frac{1+3\,S^2}{4}-\frac{3\,q\,S^2}{4}\right)^2-(S^3-q\,S^3)^2} \right)^{-\frac{1}{2}}, \\
    \vartheta_2 =& \lambda_2 . \left(\left( \frac{1+3\,S^2}{4}-\frac{3\,q\,S^2}{4}\right)^2-(S^3-q\,S^3)^2\right)^{-\frac{1}{2}}, \\
    \text{and} \hspace{0.5cm} \Gamma_2=&(\lambda_2)^3 . \left(\left( \frac{1+3\,S^2}{4}-\frac{3\,q\,S^2}{4}\right)^2-(S^3-q\,S^3)^2\right)^{-\frac{1}{2}}.
\end{align}

\subsection{Evolution of quantum criteria under the phase flip channel}
The quantum Fisher information expression for the three-qubit GHZ state under the influence of this channel is given by
\begin{equation}
    \mathbf{\textit{F}}\left(\varrho^{PFC} \right) = \frac{3}{4q}+ \frac{9}{4(4-3\,q)}  +  \frac{4}{4-3\,q} \left[\frac{\left((-12+9\,q)/16+(2S^6(1-q)\right)^2}{(4-3\,q)^2/16-(2S-1)^6(1-q)^2} - \left(\frac{9}{16}-(2S-1)^6\right)\right] .
\end{equation}
Here, the concurrence entanglement in the GHZ state under the phase-flip channel evolved as follows
\begin{equation}
    \mathbf{C}\left(\varrho^{PFC}\right) = \text{max} \left\{0, (2S-1)^3+q\left(\frac{1}{4}-6\,S+12\,S^2-8\,S^3\right) \right\}.
\end{equation}
Then, the expression for the skew information of the GHZ state under the phase-flip channel can be determined as follows
\begin{align}
    &\mathtt{I}\left(\varrho^{PFC}\right) = 8 \Bigg[ \left(\frac{3\,\vartheta_3}{16}(2\,S-1)^3(1-q)-\frac{\lambda_3}{2}(2\,S-1)^3-\frac{\Gamma_3}{4}(2\,S-1)^9(1-q)^2\right)^2 +\frac{3}{32\,q}\nonumber\\
    & +  \frac{1}{16\,\left((4-3\,q)^2/16 - (2\,S-1)^6(1-q)^2\right)}\Bigg(\frac{-3\,\sqrt{(4-3\,q)/4+\sqrt{(4-3\,q)^2/16 - (2\,S-1)^6(1-q)^2}}}{4}\nonumber \\
    &+ \frac{(2\,S-1)^2(1-q)}{\sqrt{(4-3\,q)/4+\sqrt{(4-3\,q)^2/16 - (2\,S-1)^6(1-q)^2}}}\Bigg)^2 \Bigg],  
\end{align}
where 
\begin{align}
    &\vartheta_3 = \lambda_3 \, \left(\frac{(4-3\,q)^2}{16}-(2\,S-1)^6(1-q)^2\right)^{-\frac{1}{2}}, \hspace{0.5cm}\Gamma_3  = (\lambda_3)^3 \, \left(\frac{(4-3\,q)^2}{16}-(2\,S-1)^6(1-q)^2\right)^{-\frac{1}{2}}, \\
    & \text{and} \hspace{0.5cm} \lambda_3 = \left((4-3\,q)/4+\sqrt{(4-3\,q)^2/16 - (2\,S-1)^6(1-q)^2} \right)^{-\frac{1}{2}}.
\end{align}
\begin{figure}[htbp]
		\begin{minipage}[b]{0.33\linewidth}
			\centering
			\includegraphics[width=\linewidth]{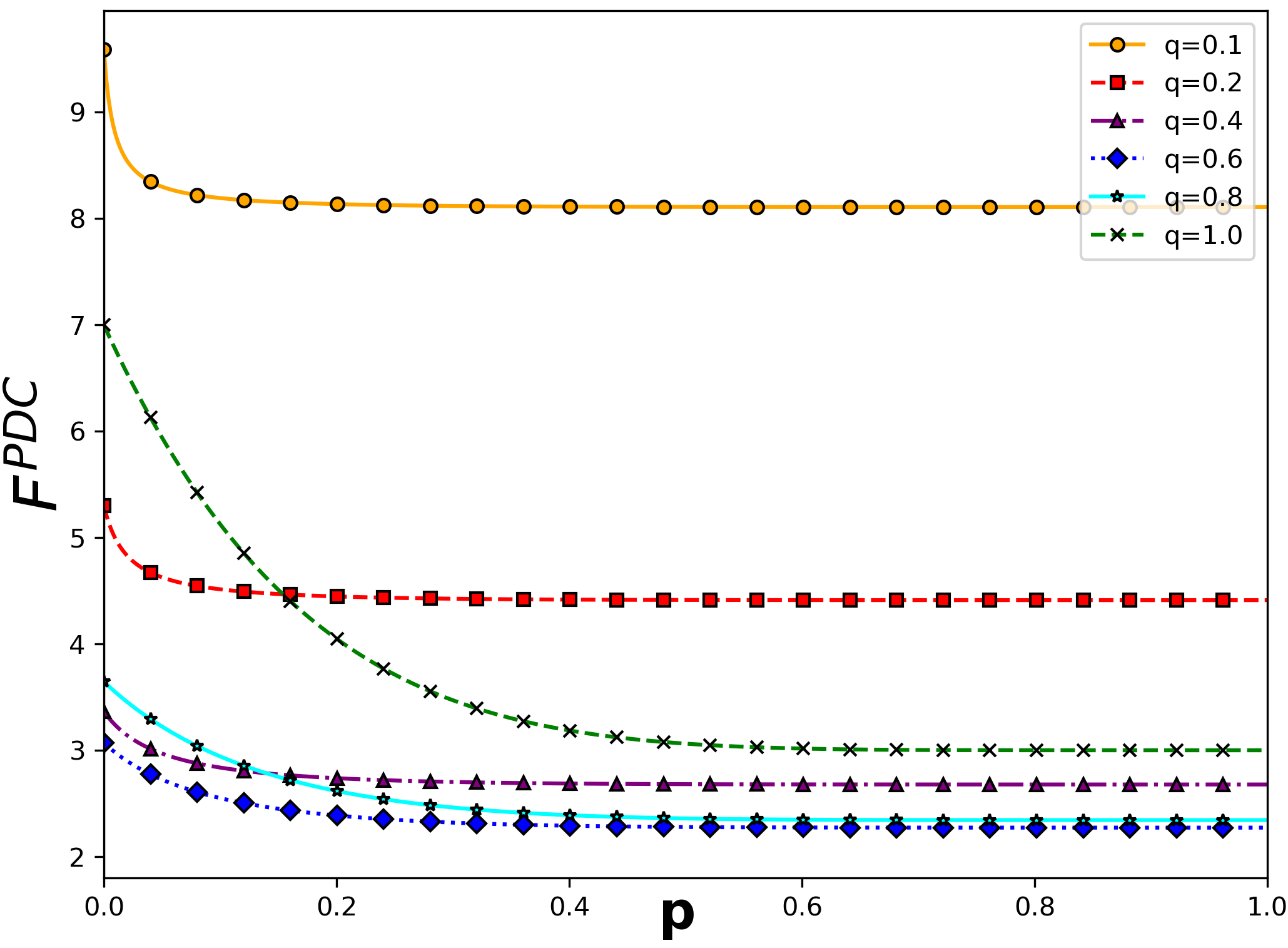}
   (I)
		\end{minipage}%
		\begin{minipage}[b]{0.33\linewidth}
			\centering
			\includegraphics[width=\linewidth]{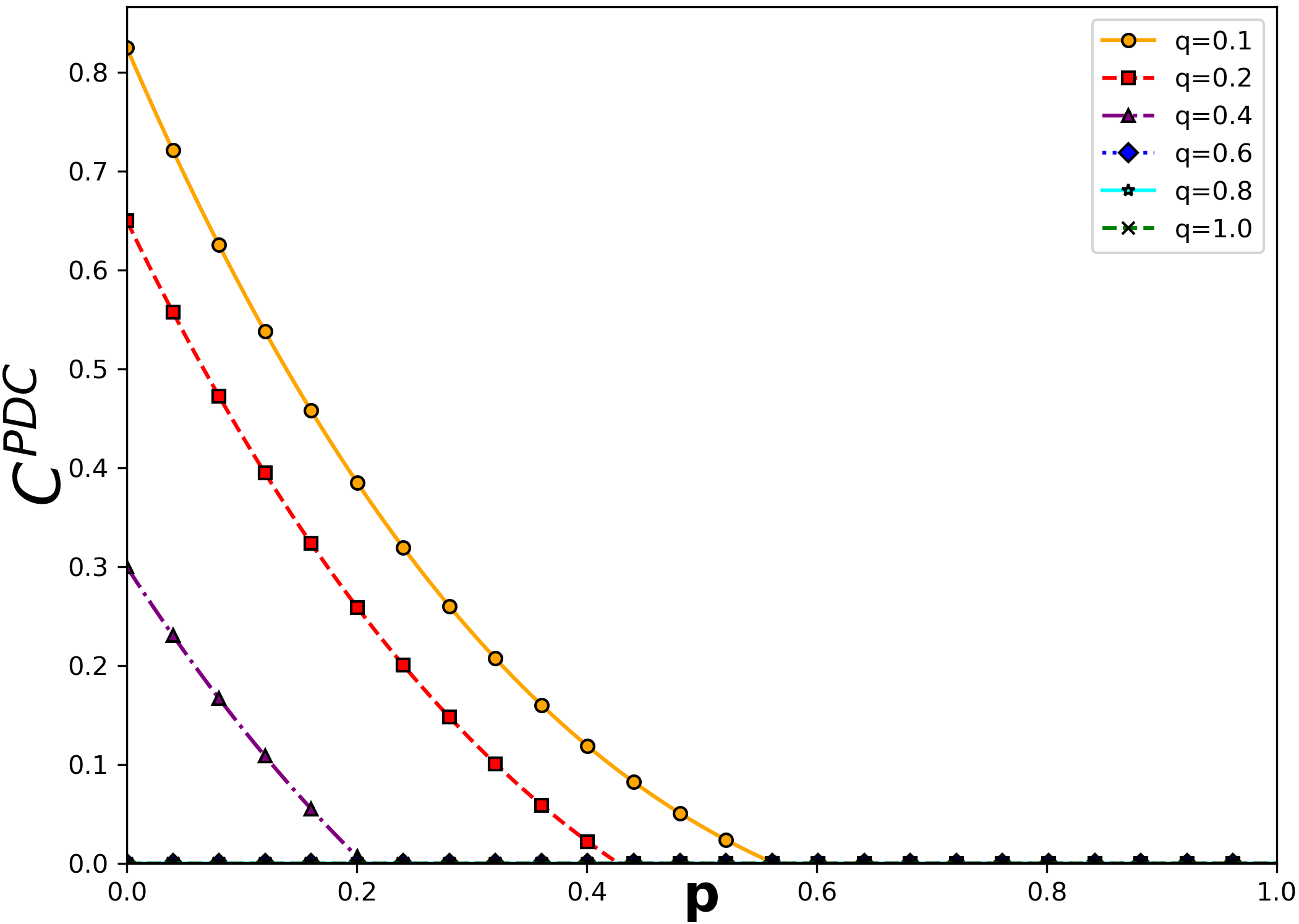}
  (II)
		\end{minipage}%
            \begin{minipage}[b]{0.33\linewidth}
			\centering
			\includegraphics[width=\linewidth]{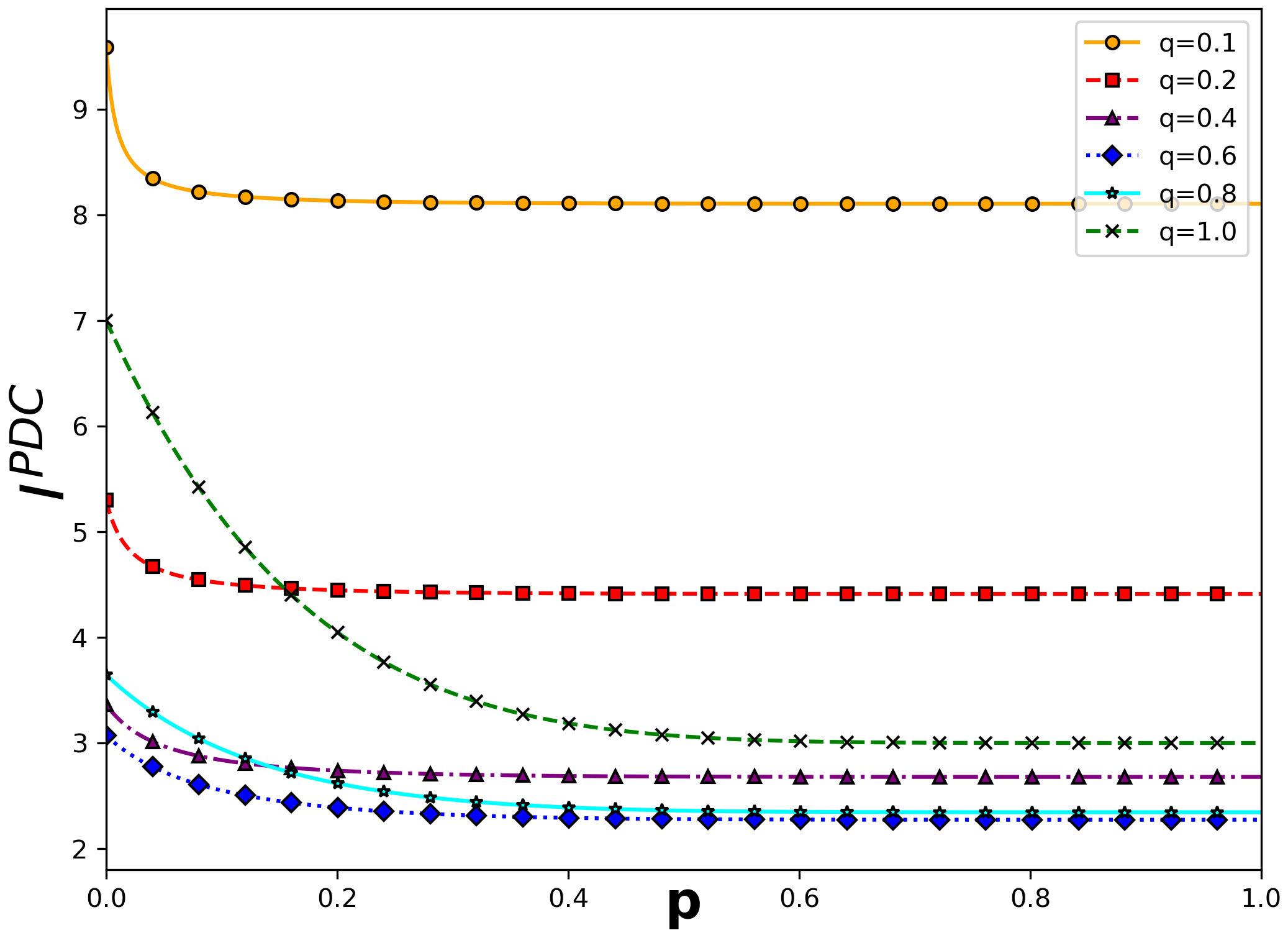}
   (III)
		\end{minipage}
		\caption{Variation of quantum Fisher information (I), Wootters concurrence (II) and skew information (III) in the phase damping channel as a function of the decoherence parameter $p$ for various values of the mixing parameter q.}\label{Fig1}
	\end{figure}
    
\begin{figure}[htbp]
		\begin{minipage}[b]{0.33\linewidth}
			\centering
			\includegraphics[width=\linewidth]{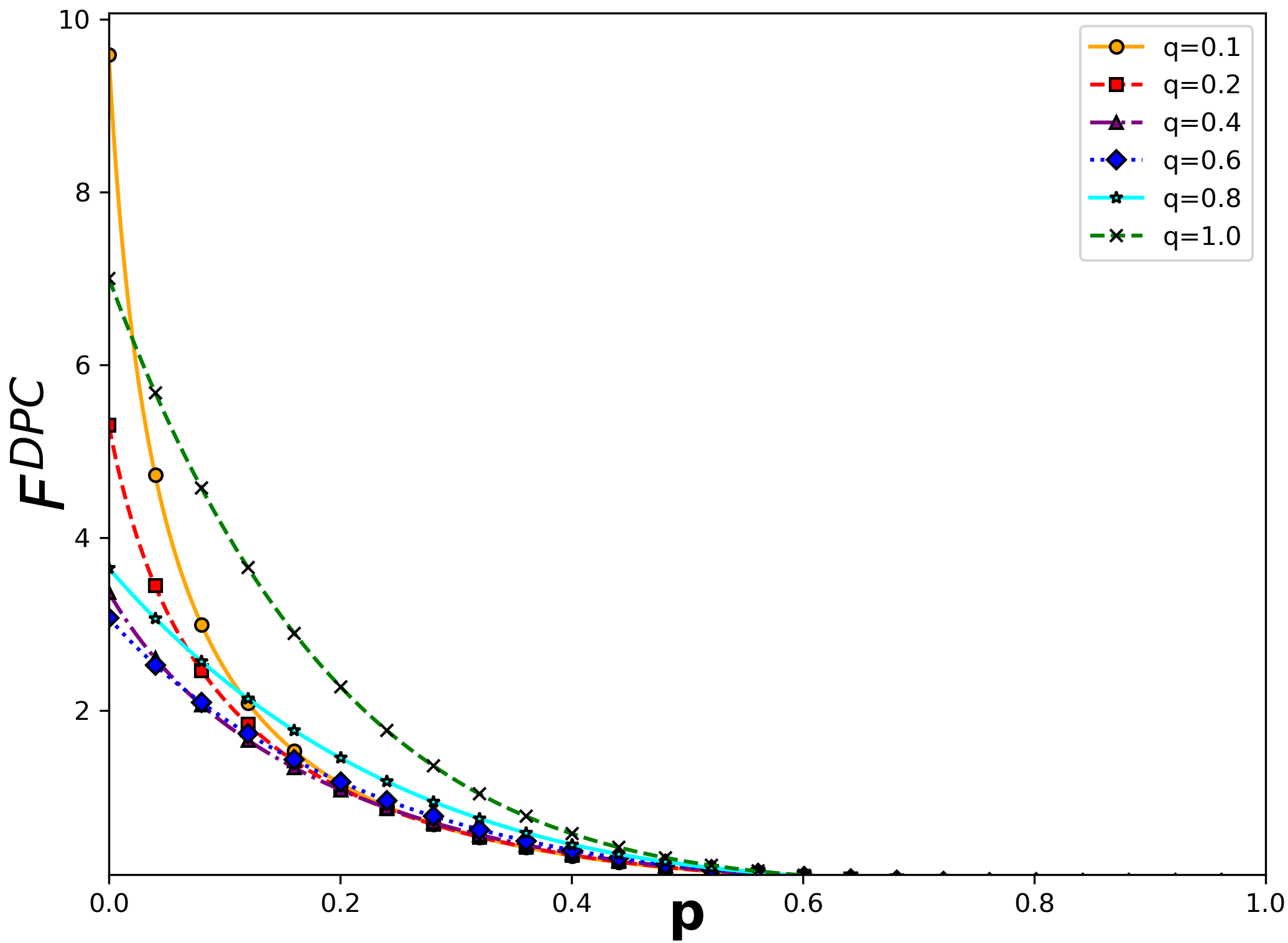}
   (I)
		\end{minipage}%
		\begin{minipage}[b]{0.33\linewidth}
			\centering
			\includegraphics[width=\linewidth]{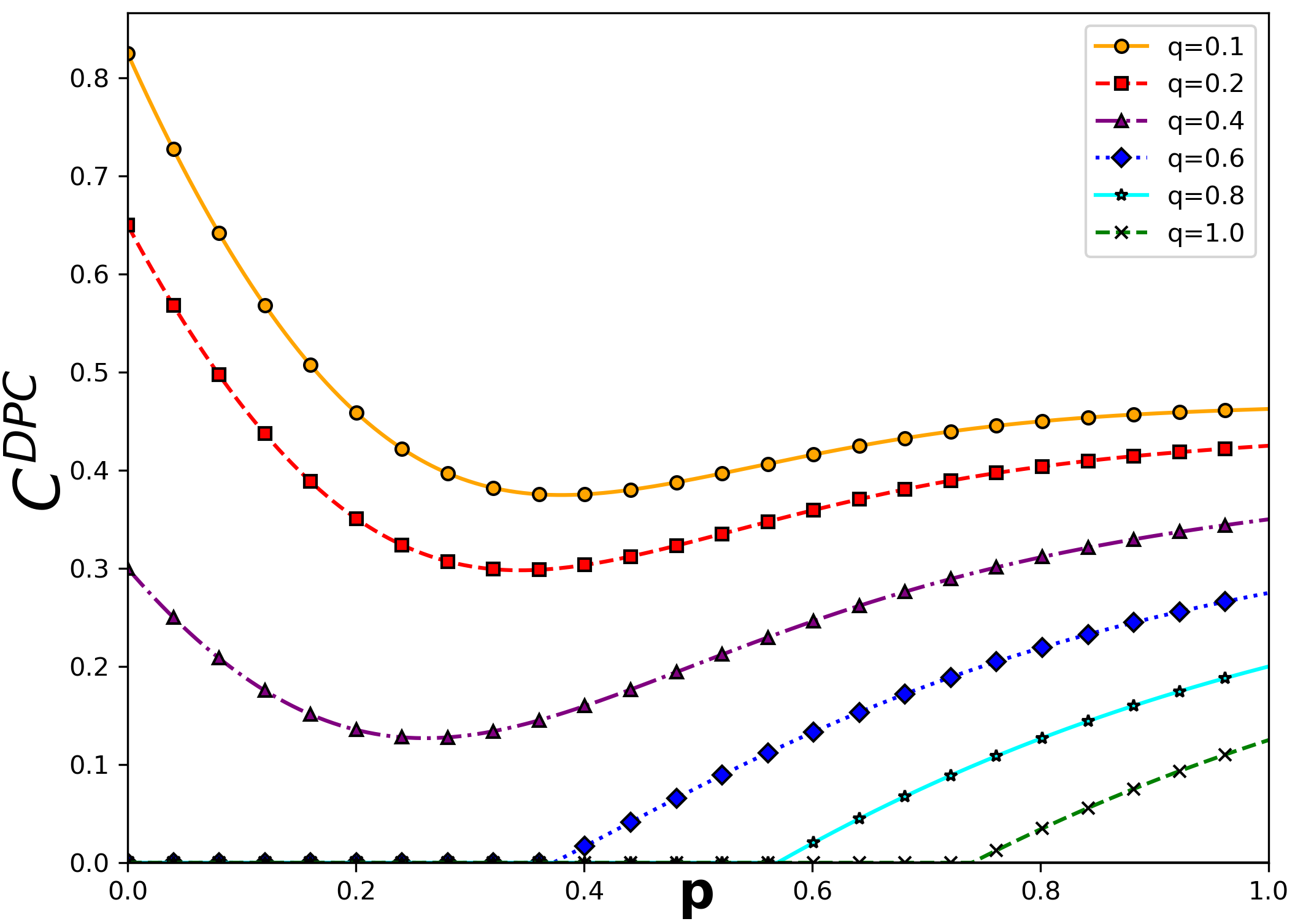}
  (II)
		\end{minipage}%
            \begin{minipage}[b]{0.33\linewidth}
			\centering
			\includegraphics[width=\linewidth]{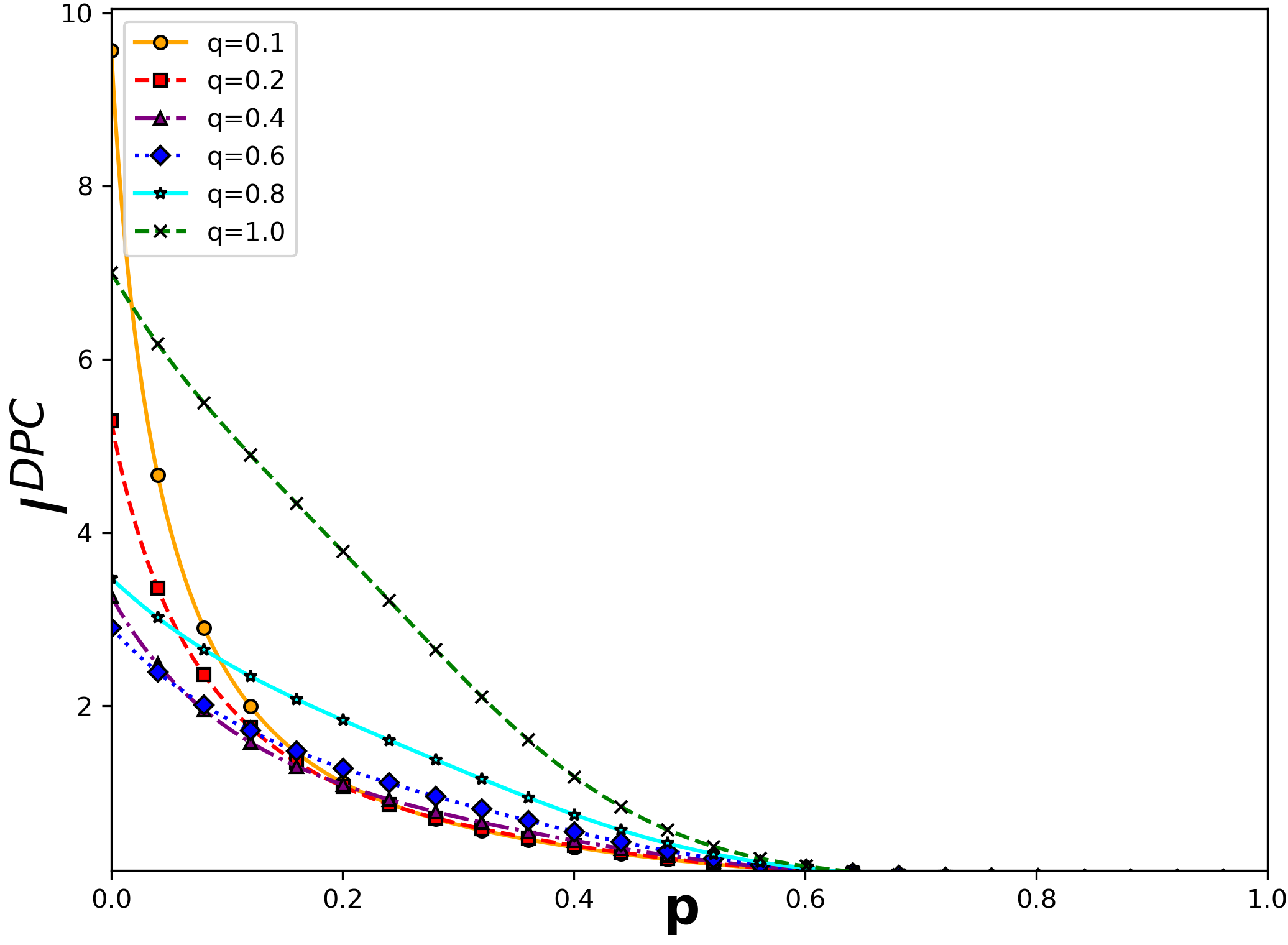}
   (III)
		\end{minipage}
		\caption {Dynamics of quantum criteria under the depolarizing channel using the same parameter sets shown in figure. \ref{Fig1}.}\label{Fig2}
	\end{figure}

\begin{figure}[htbp]
		\begin{minipage}[b]{0.33\linewidth}
			\centering
			\includegraphics[width=\linewidth]{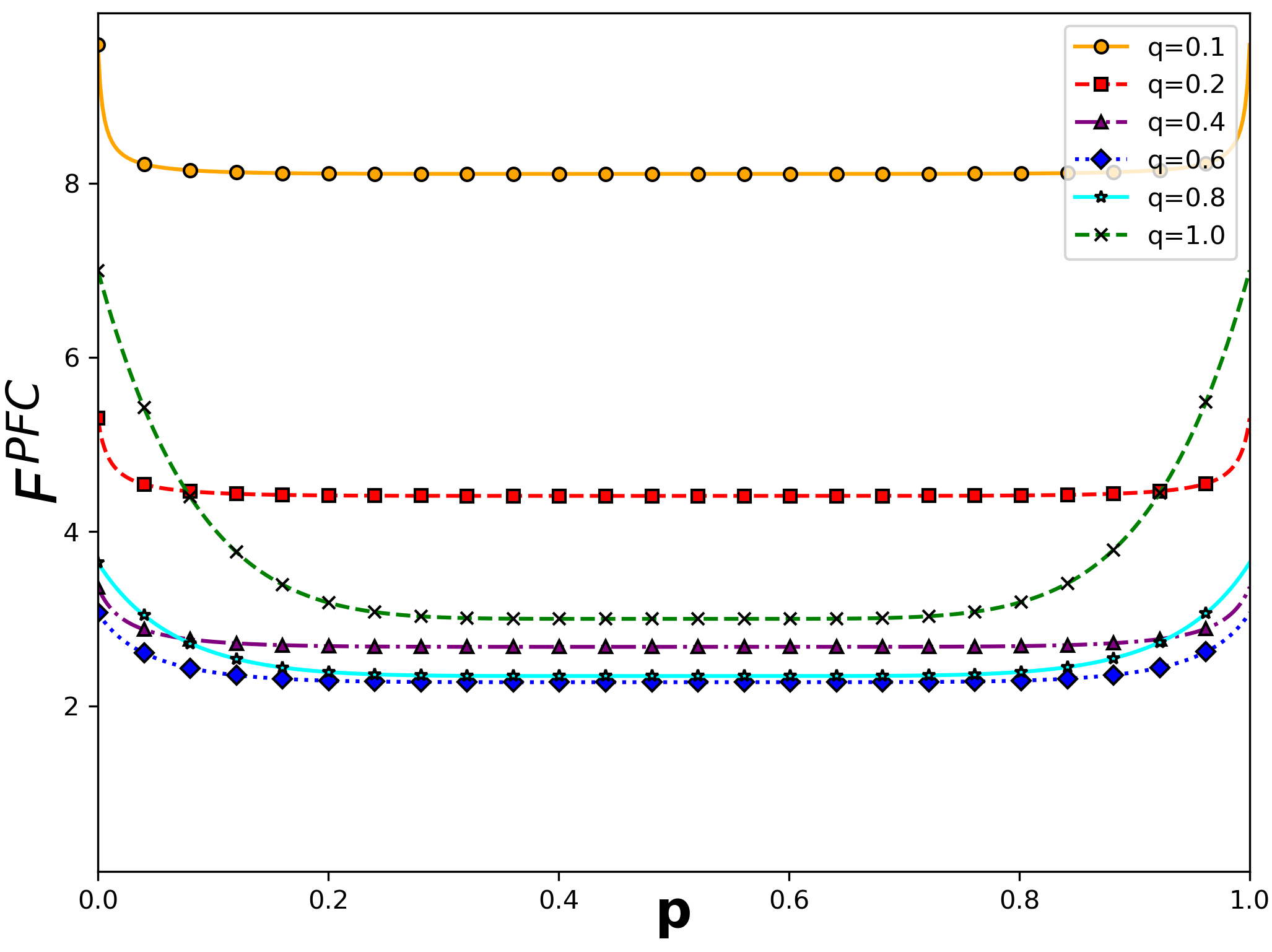}
   (I)
		\end{minipage}%
		\begin{minipage}[b]{0.33\linewidth}
			\centering
			\includegraphics[width=\linewidth]{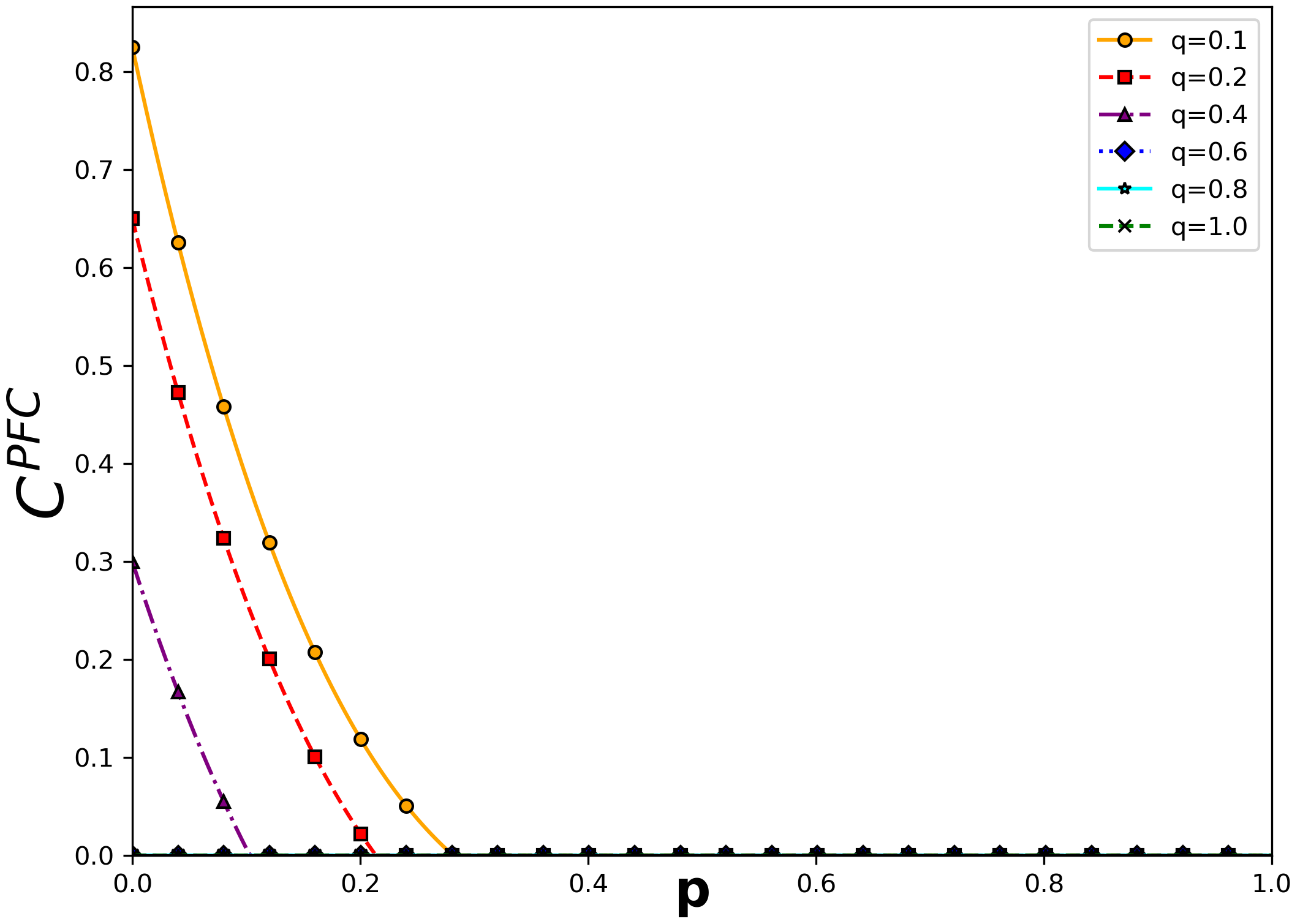}
  (II)
		\end{minipage}%
            \begin{minipage}[b]{0.33\linewidth}
			\centering
			\includegraphics[width=\linewidth]{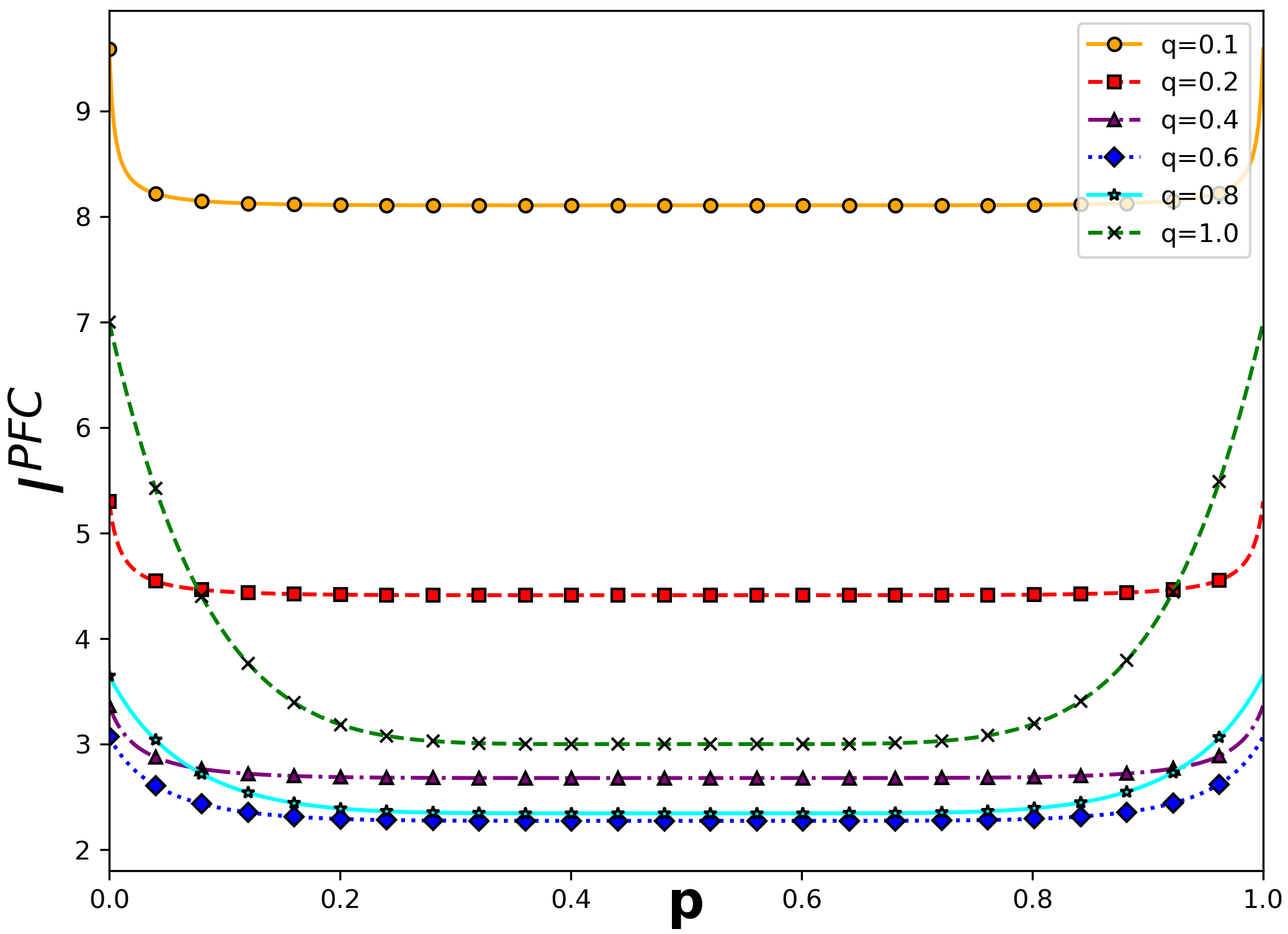}
   (III)
		\end{minipage}
		\caption {Evolutions of quantum criteria under phase flip channel using the same parameter sets shown in figure.\ref{Fig1}.}\label{Fig3}
	\end{figure}

After deriving the analytical expressions for quantum Fisher information, Wootters concurrence, and skew information under the influence of quantum decoherence, we present the dynamic behavior of these quantum criteria in Figs.\textbf{(\ref{Fig1})}, \textbf{(\ref{Fig2})}, and \textbf{(\ref{Fig3})}. The plots depict the evolution of quantum Fisher information, concurrence, and skew information for the Markovian phase-damping, depolarizing, and phase-flip channels as a function of the decoherence parameter $p$. These results demonstrate that the mixing parameter $q$ plays a significant role in shaping the dynamics of these quantum criteria throughout their temporal evolution for all the channels considered. Furthermore, the data indicate that the concurrence reaches its maximum value at small $q$ values, while the quantum Fisher information and skew information attain their highest values at the extreme ends of the parameter $q$, i.e., for both high and low q values. These observations highlight the importance of controlling the physical parameters of GHZ states in the presence of decoherence, as it can significantly affect the performance of quantum systems. Moreover, we observe that as the decoherence parameter $p$ increases, all three quantities (concurrence, quantum Fisher information, and skew information) decrease, eventually approaching zero for the depolarizing channel. In contrast, for the phase damping and phase flip channels, the quantum Fisher information and skew information decrease up to $p=0.2$, after which both quantities remain constant, even with further increases in $p$. This suggests a saturation effect in the decoherence dynamics for these channels, where beyond a certain point, further decoherence does not significantly affect the quantum properties under study. These findings underscore the complex interplay between the decoherence parameter and the channel-specific effects on quantum information measures, providing insights into the resilience and decay of quantum resources in noisy environments.

    
On the other hand, the graphs show that concurrence decreases more rapidly than quantum Fisher information and skew information, which decrease more gradually. This suggests that concurrence undergoes a much sharper decline under decoherence, while quantum Fisher and skew information exhibit a more moderate reduction. Notably, noise in the phase damping and phase flip channels less affects these quantities than in the depolarizing channel, suggesting that the type of decoherence channel significantly influences the behavior of these quantum information measures. Moreover, quantum Fisher information and skew information display similar behavior across the three decoherence channels, further emphasizing their comparable roles in quantum estimation. Interestingly, both Fisher information and skew information reach their maximum values when entanglement, as measured by concurrence, is strongest. This suggests that GHZ states provide the most accurate estimation of unknown parameters when entanglement is preserved, reinforcing the idea that entanglement plays a crucial role in quantum sensing and estimation. These findings also highlight an important point: skew information behaves similarly to quantum Fisher information in quantum estimation theory. Both quantities provide critical insights into the precision of parameter estimation in quantum systems. Therefore, when using concurrence or entanglement as a reference for quantum resource quantification, skew information can effectively replace quantum Fisher information for assessing estimation accuracy. In summary, skew information is not only a viable substitute for quantum Fisher information in the context of quantum estimation theory; it also complements Fisher information by offering additional, relevant insights into the system's dynamics.

\section{Conclusion}
In this work, we first studied in detail the skew information and quantum Fisher information, defined using the symmetric logarithmic derivative (SLD), for a three-qubit system subjected to decoherence. Both measures are important tools for quantifying various quantum resources, including quantum coherence \cite{Yu2017}, asymmetry \cite{Sun2021}, and quantum correlations \cite{Slaoui12019}. We specifically focused on the evolution of these two quantities for the three-qubit GHZ state under the phase damping, depolarization, and phase flip channels. We then compared the results from these measurements with the Wootters concurrence, a widely recognized measure for assessing quantum entanglement. Our research showed that quantum Fisher information and skew information behave similarly when decoherence happens, having similar properties and dynamics across the various noise channels. This implies that both measures serve as reliable indicators of quantum resources and can function interchangeably in specific situations, particularly when assessing the accuracy of quantum parameter estimation. Furthermore, our study demonstrated that quantum correlations play a critical role in improving the accuracy and reducing errors in quantum parameter estimation. In particular, we observed that the maximum quantum Fisher information coincides with the maximum entanglement in the GHZ state, highlighting the close relationship between entanglement and the efficiency of quantum metrology. The analysis of skew information further supported these findings, confirming the central role of quantum entanglement in enhancing the performance of quantum estimation protocols.\par 

In summary, this work highlights the significance of both quantum Fisher information and skew information in comprehending and optimizing quantum systems exposed to decoherence. It provides analytical solutions for any three-qubit state and underscores the crucial role of quantum entanglement in maximizing the precision of quantum metrology. These measures are invaluable in the development of quantum technologies.


\end{document}